\documentclass[article]{aa}  
\pdfoutput=1
\bibpunct{(}{)}{;}{a}{}{,} 
\usepackage[T1]{fontenc}
\usepackage{textcomp}
\usepackage{graphicx}
\usepackage{multirow}
\usepackage{pifont}
\usepackage{makecell} 
\setcellgapes{5pt}
\usepackage{txfonts}
\usepackage{rotating}
\usepackage[]{xcolor}
\usepackage{comment}
\usepackage{amsmath,bm}
\usepackage{booktabs}
\usepackage{nicematrix}
\usepackage{hyperref}

\newcommand{\hii}{\mbox{H\,{\sc{ii}}\,}}

\usepackage{graphicx}
\usepackage{txfonts}
\usepackage{tikz}
\usetikzlibrary{calc,positioning}
\usepackage{ulem}
\usepackage{subfigure}
\usepackage{ifthen}
\usepackage{placeins}
\usepackage{lipsum}
\usepackage{float}
\usepackage{xcolor}
\usepackage{xspace}
\usepackage{arydshln}

\hypersetup{colorlinks=true,linkcolor=blue,citecolor=blue,filecolor=blue,urlcolor=blue}

\newcommand{\kms}{\mathrm{km\,s^{-1}}}

\newcommand{\pc}{\mathrm{pc}}
\newcommand{\MSun}{\mathrm{M}_\odot}
\newcommand{\kelvin}{\mathrm{K}}
\newcommand{\Myr}{\mathrm{Myr}}

\newcommand{\ndens}{\mathrm{cm^{-3}}}

\newcommand{\coldensMPC}{\MSun\,\pc^{-2}}

\newcommand{\eff}{\varepsilon_\mathrm{ff}}

\newcommand{\simuSFR}{\dot{M}_\star}
\newcommand{\tff}{t_\mathrm{ff}}

\newboolean{ActivateComments}
\setboolean{ActivateComments}{true}

\newboolean{ActivatePaolosComments}
\setboolean{ActivatePaolosComments}{false}

\newboolean{RefComments}
\setboolean{RefComments}{true}

\makeatletter
\newcommand*\bigcdot{\mathpalette\bigcdot@{.5}}
\newcommand*\bigcdot@[2]{\mathbin{\vcenter{\hbox{\scalebox{#2}{$\m@th#1\bullet$}}}}}
\makeatother

\DeclareRobustCommand{\Pcomment}[2]{%
  \ifthenelse{\boolean{ActivatePaolosComments}}{%
    \ifthenelse{\equal{#1}{}}{\textcolor{cyan}{#2}}{\textcolor{gray}{\sout{#1}}\textcolor{cyan}{#2}}}{#2}}

\newcommand{\Refcomment}[1]{%
  \ifbool{RefComments}{\textbf{#1}}{}%
}

\newcommand{\Acomment}[1]{%
  \ifbool{ActivateComments}{\textcolor{cyan}{#1}}{}%
}

\newcommand{\ANcomment}[1]{%
  \ifbool{ActivateComments}{\textcolor{green}{#1}}{}%
}

\newcommand{\ATcomment}[1]{%
  \ifbool{ActivateComments}{\textcolor{red}{#1}}{}%
}

\definecolor{lime}{HTML}{A6CE39}
\DeclareRobustCommand{\orcidicon}{%
        \begin{tikzpicture}
        \draw[lime, fill=lime] (0,0) 
        circle [radius=0.16] 
        node[white] {{\fontfamily{qag}\selectfont \tiny ID}};
        \draw[white, fill=white] (-0.0625,0.095) 
        circle [radius=0.007];
        \end{tikzpicture}
        \hspace{-2mm}
}

\newcommand{\orcidPS}{\href{https://orcid.org/0000-0001-7044-3809}{\orcidicon}}

\newcommand{\lowturb}{\texttt{n1f3}\xspace}
\newcommand{\highturb}{\texttt{n1f10}\xspace}
\newcommand{\highn}{\texttt{n1.5f10}\xspace}
\newcommand{\refsim}{\texttt{n1f7}\xspace}
\newcommand{\outflowsim}{\texttt{n1f7outf}\xspace}

\defcitealias{Hennebelle2026}{Paper~I}

\begin{document} 
   \title{What sets the star-formation rate in the Milky Way?}
   \subtitle{II. Comparing the star-forming clumps from simulations and observations}

   \author{P.~Suin
          \inst{1}\thanks{\email{paolo.suin@cea.fr}}\orcidPS
          \and
          P.~Hennebelle\inst{1}
          \and 
          G.~Abboudeh\inst{1}
          \and
          D.~Elia\inst{2}
          \and
          S.~Molinari\inst{2, 3}
          \and
          E.~Schisano\inst{2, 3}
          \and 
          N.~Brucy\inst{4}
          \and
          S.~Glover\inst{5, 6}
          \and
          U.~Lebreuilly\inst{1}
          \and
          S.~Mercimek\inst{7}
          \and
          A.~Nucara\inst{2, 3}
          \and
          V.-M.~Pelkonen\inst{2}
          \and
          J.~D.~Soler\inst{8}
          \and
          A.~Traficante\inst{2}
          \and
          R.~S.~Klessen\inst{5, 6}
          \and
          L.~Testi\inst{9}
           } 
    \institute{
    Université Paris-Saclay, Université Paris Cité, CEA, CNRS, AIM, F-91191 Gif-sur-Yvette, France
    \and
    INAF -- Istituto di Astrofisica e Planetologia Spaziali, Via Fosso del Cavaliere 100, I-00133 Roma, Italy
    \and
    Dipartimento di Fisica, Università di Roma Tor Vergata, Via della Ricerca Scientifica 1, 00133 Roma, Italy
    \and
    ENS de Lyon, CRAL UMR5574, Université Claude Bernard Lyon 1, CNRS, Lyon 69007, France
    \and
        Universität Heidelberg, Zentrum für Astronomie, Institut für Theoretische Astrophysik, Albert-Ueberle-Str. 2, 69120 Heidelberg, Germany
    \and
    Universität Heidelberg, Interdisziplinäres Zentrum für Wissenschaftliches Rechnen, Im Neuenheimer Feld 225, 69120 Heidelberg, Germany
    \and
    Jodrell Bank Centre for Astrophysics, UK ALMA Regional Centre Node, University of Manchester, Manchester M13 9PL, UK
    \and
    University of Vienna, Department of Astrophysics, Türkenschanzstraße 17, 1180 Wien, Austria
    \and
    Università di Bologna, Dipartimento di Fisica e Astronomia `Augusto Righi', Via Gobetti 93/2, 40129, Bologna, Italy
    }
  \date{Received: 15 June 2026 / Accepted: 18 October 2026}
  \abstract
    {Linking the star formation (SF) properties of dense clumps to their parent large-scale environment is a key step to interpret Galactic surveys. Achieving a consistent, multi-scale picture of SF requires a comparative analysis of observations and simulations.}
   {We exploit kiloparsec-scale magneto-hydrodynamic simulations designed to reproduce local Milky Way conditions. 
   Leveraging the wide range of scales covered by the simulations (reaching a resolution of $0.03\,\pc$), we aim to compare the properties of the simulated population of star-forming clumps with observational catalogues.}
   {After extracting the star-forming structures with the \texttt{HOP} clump-finding algorithm, we compare their physical (mass and size) and SF properties (SF rate and efficiency) with those of the Hi-GAL compact source catalogue.}
   {Despite the different methodologies and the three-dimensional nature of our clump extraction, the simulated clumps show good agreement with the observed sample in their mass and size distributions. We discover a consistently linear relation between mass and SF rate (SFR) for the simulated clumps, which is remarkably independent of the initial conditions of the boxes. Using scaling relations to distinguish between turbulence- and gravity-regulated gas inflow onto the clump scales, we find tentative evidence in favour of a gravity-regulated regime of SF at the probed sub-parsec scales. 
   }
   {}

   \maketitle

\section{Introduction}
A strong relationship between the mass of the gas reservoir and local star formation (SF) activity is observed across a wide range of spatial scales. The best-known example is the Kennicutt--Schmidt relation \citep{Schmidt1959, Kennicutt1989}, linking the gas column density to the SF rate (SFR) per unit area on galactic scales. However, later studies found that focusing on molecular gas provided better agreement \citep{Bigiel2008}, with a nearly linear relationship from kiloparsec scales down to cloud scales (tens of parsecs; \citealt{Bigiel2008, Lada2010, Evans2014}). This coupling remains and even tightens when considering only the dense gas above $10^4\,\ndens$ (typically traced with HCN ($1-0$) line; \citealt{Gao2004, Wu2005, Garcia-Burillo2012, Liu2016, Stephens2016, Jimenez-Donaire2019, Neumann2025}). This dense component is mainly stored in clumps, compact (between $0.1$ and $1\,\pc$ in radius) and cold ($\lesssim 20\,\mathrm{K}$) structures, which can be identified mostly thanks to their mid- and near-infrared emission \citep{Traficante2015, Elia2017, Urquhart2018}. This makes clump populations an ideal laboratory to test SF laws on sub-parsec scales.

These empirical SF laws are useful for simplifying the description of this highly multi-scale, multi-physics process. However, SF is not a sequence of independent steps occurring at well-separated scales, so the physical origin of these relations is difficult to infer from a simple power-law link between gas mass and SF activity alone \citep{Heiderman2010, Federrath2012, Kennicutt2012, Federrath2013, Krumholz2014}. The interstellar medium naturally organises into a hierarchy of nested structures (clouds, clumps, cores) that are continuously assembled, destroyed and fed by the larger-scale environment \citep{Klessen2016, Girichidis2020}. 
This coupling creates a sort of environmental memory that connects the instantaneous properties of star-forming clumps (such as mass, density distribution and accretion rate) to the properties and past history of the medium that created them \citep[see, for example][]{Jimenez-Donaire2019, Neumann2025}.

Therefore, understanding this connection between scales (and between the star-forming dense clumps and the interstellar medium) often requires large statistical samples to cover the large parameter space of the physical quantities involved. Galactic far-infrared and submillimetre surveys such as ATLASGAL \citep{Schuller2009, Urquhart2014, Urquhart2022}, the JCMT Plane Survey \citep{Eden2017, Eden2024}, and the Herschel infrared Galactic Plane Survey (Hi-GAL; \citealt{Molinari2010, Traficante2015, Molinari2016, Elia2017, Elia2021}) provide a large statistical sample on which SF laws can be tested in a variety of environments. In particular, the Hi-GAL survey produced a homogeneous census of compact structures across the full Galactic plane, identifying more than $150\,000$ sources along with their masses, sizes, evolutionary stage, and distance estimates \citep{Elia2017, Elia2021}. This allowed the construction of a complete map of the Galactic SF activity at sub-kpc resolution \citep{Elia2025}.

Observational biases, however, complicate the interpretation of the data. Projection effects and distance uncertainties can affect the derived physical properties (e.g., \citealt{Baldeschi2017, Elia2017, Urquhart2018, Elia2021, Lebreuilly2025}). In addition, SFR estimates generally rely on tracers or theoretical models that naturally average the SFR over a window comparable to the clumps' lifetimes \citep[order of one to a few Myr, see][]{Battersby2017, Urquhart2022}. For example, the Hi-GAL clumps lack direct measurements of their internal SFR, so \citet{Elia2025} resorted to semi-analytical methods and evolutionary tracks to infer them (see also \citealt{Veneziani2017}). Although this approach proved robust when averaging over the entire ensemble \citep{Elia2022}, by construction it does not recover the actual SF properties of the clumps, so the effect of the larger-scale environment cannot be probed.

Numerical simulations can provide a complementary approach. They allow us to follow the system's temporal evolution while maintaining the local three-dimensional quantities accessible.
Direct measurements of the instantaneous SFR are straightforward (e.g., via sink particle accretion), and the controlled environment allows disentangling the effects of global parameters on the final SF process. Nevertheless, computational costs and the complexity of all the physical processes involved in the multi-scale SF process require compromises. 

Several studies have used simulations of isolated star-forming clouds to study their fragmentation and SF process at smaller scales ($0.1-10\,\pc$; see, for example, \citealt{Smith2008, Gomez2014, Grudic2021, Priestley2026}). However, the starting setup generally features idealised clouds, with well-defined initial turbulence and density. It follows that the parameter space of the environmental properties is generally quite limited, and the impact of the outer environment is, by definition, neglected. Zoom-in simulations may help in this regard, creating a more realistic starting point, but the population of substructures remains limited \citep{Seifried2017, Hennebelle2018, Smith2020, Feng2024}.
Alternatively, simulations can feature large domains, capable of reproducing the Galactic disc or portions of it (e.g., \citealt{Sormani2020, Tress2020, Tress2021, Keller2022, Khullar2024, Renaud2024,  Colman2024, Colman2025, Goller2025}). In these, one can recover the formation of molecular clouds, but requires subgrid models to recover the SF at smaller scales. This may create a perfectly self-consistent environment that, in practice, does not represent realistic scenarios observed. Therefore, continuous communication between the two approaches is needed to assess their limitations and advance in understanding of the SF process. 

In this context, the intermediate regime between kpc-scale environments and sub-parsec star-forming structures remains much less studied, leaving the connection between the SF process within star-forming clumps and the large-scale environment underexplored. 
The simulation suite presented in the companion paper \citep[][hereafter \citetalias{Hennebelle2026}]{Hennebelle2026} was designed to fill this gap by reproducing local Galactic disc on kiloparsec scales while achieving sub-parsec resolution. 
In this work, we study the emergence of a sub-parsec-scale SF law from a simulated population of clumps and compare the properties of the simulated structures with those of the observed Hi-GAL compact sources. Exploiting the large dynamic range of the simulation set, we are able to extract a statistically relevant sample of star-forming clumps arising within realistic clouds. Through a direct analysis, we aim to test whether the simulated clump populations can reproduce the observed distributions in mass and size, and to characterise how large-scale properties influence clump-scale SF laws. 

The paper is organised as follows. Section~\ref{sec: methods} describes the simulations, the selected subsample of Hi-GAL clumps, and the clump extraction procedure from the numerical domain. In Section~\ref{sec: results}, we present a comparison between the observed and simulated populations, including SF relations, and discuss the results in Section~\ref{sec: discussion}. Finally, we summarise our conclusions in Section~\ref{sec: conclusions}.

\section{Methodology}
\label{sec: methods}
\subsection{Simulation suite}
The five simulations analysed here are described in \citetalias{Hennebelle2026}. For the sake of completeness, we summarise the most important characteristics. The magneto-hydrodynamics simulations were performed with the \texttt{RAMSES} adaptive mesh refinement (AMR) code \citep{Teyssier2002}. The domain consists of a kpc box with periodic boundaries in $x$ and $y$ and open boundaries in the vertical direction $z$. The grid starts at $256^3$ and is refined wherever the local Jeans length, $\lambda_\mathrm{J}$, falls below 10 resolution elements. The AMR grid reaches up to fifteen levels of refinement, corresponding to a maximum resolution of $\Delta x\approx 0.03\,\pc$.
The initial setup has been constructed to resemble the average conditions observed in the Milky Way. The density distribution follows a standard Gaussian disk profile, uniform across the $x$ and $y$ axes and decreasing along the $ z$-axis. The disk is thus characterised by a mid-plane density, $n_0$, and a scale height of $150\,\pc$. 
The magnetic field is initially uniform, with an intensity of $3.5\,\mu\mathrm{G}$ and oriented perpendicularly to the $xy$ plane.

To let the disk relax from the initial conditions, turbulence is continuously driven in the $x$ and $y$ directions (see also \citealt{Brucy2023} and \citetalias{Hennebelle2026}), and applied at low altitude ($\leq75\,\pc$). The turbulence driving consists of an equal mixture of compressive and solenoidal modes (being two-dimensional), the strength of which is controlled by the $f_\mathrm{rms}$ parameter. We varied $f_\mathrm{rms}$ between 3000, 7000, and 10\,000 (in code units, see also Table~\ref{tab: simulations set}). These forcing lead to 1D velocity dispersion on the disc plane of the order of $9-15\,\kms$ ($\approx9\,\kms$ with the smallest $f_\mathrm{rms}$, and $15\,\kms$ with the highest). During the initialisation phase, the grid refinement is switched off.

After $\approx150-200\,\Myr$ (roughly two turbulent crossing times), the turbulence is fully developed, and the flow reaches a steady state. Before allowing the grid to refine up to the highest AMR level, we inject supernovae into the medium to restructure the environment, while keeping the turbulence driving active. Letting supernovae set off in the highest-density cells at a rate comparable to the Milky-Way SFR, these disperse the densest regions that would otherwise undergo immediate collapse if allowed to refine. We then evolve the simulations for a few Myr at an intermediate resolution (allowing refinement up to level 11; $\Delta x\approx0.5\,\pc$), to let the system rebuild the high-density tail of the gas density probability distribution function. We refer to \citetalias{Hennebelle2026} for a detailed description of the methodology and its motivation.
Finally, we allow refinement up to the maximum level and activate sink-particle formation in the densest collapsing regions \citep{Bleuler2014}. We adopted a sink formation threshold of $n_\mathrm{sink}=2\times10^5\,\ndens$, roughly corresponding to a local $\lambda_\mathrm{J}$ of $2\Delta x$ for a medium at $10\,\kelvin$.  

The simulations also include pre-supernova feedback. Every time $100\,\MSun$ of gas is converted into sinks, we spawn a stellar particle in the sink that has accreted the most since the previous stellar-particle formation event. Its mass is randomly sampled from a \citet{Salpeter1955}, restricted to $m_\star\ge 8\,\MSun$. Then, the hosting sink begins emitting ionising photons according to the prescription of \citet{Colling2018}, thereby generating an \hii region.

In one of the simulations (\outflowsim), protostellar outflows are implemented \citep{Verliat2022}. As a sink accretes material from the surroundings, 1/3 of it is ejected along the angular momentum axis of the sink, consistent with previous observational \citep{Bacciotti2011, Nisini2018} and numerical results \citep{Hennebelle2008, Matsushita2017}. The material is launched at a fraction of the sink's escape velocity, typically resulting in $\sim30\,\kms$.

\begin{table}
\renewcommand{\arraystretch}{1.1} 
\caption{Parameter space covered by the simulation suite.}
\label{tab: simulations set}
\setlength{\tabcolsep}{5pt} 
  \centering
  \begin{tabular}{c c c c}
    \hline\hline
    Name & $\Sigma_0\,[\coldensMPC]$ & $f_\mathrm{rms}$ & $\sigma\,[\kms]$\\
    \hline\hline
    \refsim  & 12 & 7000 & 9 \\
    \lowturb  & 12 & 3000 & 7.5 \\
    \highturb  & 12 & 10\,000 & 12 \\
    \highn  & 19 & 10\,000 & 12 \\
    \outflowsim  & 12 & 7000 & 9 \\
    \hline\hline
  \end{tabular}
  \vspace{0.2cm}
\end{table}

In total, we use five simulations with varying initial mid-plane density and turbulence driving, $f_\mathrm{rms}$. 
We name the simulations by their initial midplane number density, $n_0$, and their forcing $f_\mathrm{rms}/1000$. Therefore, we denote by \refsim the simulation with forcing $f_\mathrm{rms}=7000$ and $n_0=1\,\ndens$, and by \outflowsim its sibling run that includes outflows. \lowturb and \highturb indicate the simulations having the $n_0=1\,\ndens$ and a weaker or stronger driver, respectively. Finally, we use \highn for the simulation with higher turbulence and $n_0=1.5\,\ndens$ (and $\Sigma=19.1\,\MSun\,\pc^{-2}$).

\subsection{Hi-GAL sources}
The distinction between cores, clumps, and clouds is not sharp, and these categories blend into one another. Consequently, it is often challenging to extrapolate results from numerical simulations to observational analyses, and a thorough comparison requires a direct matching between the recovered structures. To ensure that the structures identified in our simulations are compatible with those detected in observations, we use the Hi-GAL compact sources catalogue as our reference sample (\citealt{Elia2021}; see also \citealt{Molinari2016, Elia2017}) and tune the parameters of our clump-finding algorithm so that the simulated clumps have the same characteristic spatial scale.
This strategy enables a two-way approach that allows us to both test the robustness of the extraction methodology and interpret the observed (star-forming) clump properties, thanks to the local physical information available in the simulations.

The whole sample consists of 150223 structures, whose physical quantities range across several orders of magnitude -- six in mass (from $0.1\,\MSun$ to $10^5\,\MSun$) and three in size (from $\approx10^{-3}$ to a few pc). This wide variety mainly arises from the survey's extensive coverage (spanning the entire Galaxy). Recovered sources are located at different distances (from a few hundred pc to $15\,\mathrm{kpc}$), and consequently have different physical resolutions. The authors differentiate the set into cores, clumps, and clouds, characterised, respectively, by radii $R<0.05\,\pc$, $0.05\leq R\leq 1.5\,\pc$ and $R>1.5\,\pc$ (see also \citealt{Zinnecker2007}). 
As we are interested in analysing the star-forming properties of these dense structures, we consider only sources associated with SF activity and with an assigned heliocentric distance.

\subsection{Retrieving the clumps}
To recover coherent density structure from the simulations, we employed \texttt{HOP}\footnote{As described in \citet{Colman2024}, the algorithm was originally designed for $N$-body simulations. However, as the algorithm estimates the density of each particle using kernel smoothing, we can adapt it to work on the AMR grid of our simulations by skipping this first passage and forcing the AMR cells to be read as particles, as \texttt{RAMSES} naturally stores their local density.} \citet{Eisenstein1998}.  
The algorithm takes three parameters as input. The background density $n_\mathrm{thr}$, which delineates the boundary of the structures; the peak density $n_\mathrm{peak}$, which sets the minimum density for a local density peak to be recognised as a separated structure; and the saddle density $n_\mathrm{saddle}$, which governs the regrouping of the structures. If the density at the interface of two neighbouring clumps exceeds $n_\mathrm{saddle}$, the two are merged.
Given the resolution of our simulation, it is important to note that the sink particles do not represent a pure stellar population, but a mixed reservoir of stars and unresolved gas. We therefore include them in the density field used by \texttt{HOP}, distributing each sink mass on a uniform sphere of radius $2\Delta x$.

\subsection{Choice of the \texttt{HOP} parameters}
Obviously, recovering the dense star-forming sources from the simulated three-dimensional domain would not yield the exact same type of structures identified in the Hi-GAL catalogue. Indeed, the \texttt{HOP} algorithm finds clumps using a precise volume-density threshold, whereas sources in \citet{Elia2021} are extracted from the five intensity maps of Hi-GAL bands. Additionally, all structures in the simulations have the same spatial resolution (aside from the precise local AMR grid structure), while some properties of Hi-GAL sources are affected by their heliocentric distance \citep{Elia2017, Elia2021}. 
Therefore, the parameter space allows some freedom in defining what a clump is for the purposes of this comparison. Lowering the background and density thresholds in \texttt{HOP} would result in larger structures composed of multiple merged peaks. We can expect the same to be true for observed clumps as the distance increases (i.e., as the spatial resolution decreases; see also \citealt{Baldeschi2017}). Vice versa, increasing the density threshold in \texttt{HOP} will recover more separated and compact structures, as if observing clumps at higher spatial resolution. 
There is no single preferred solution, as the distinction among cores, clumps, and clouds is inherently ambiguous. Both the observational and numerical definitions are sensitive to the parameters used (see Appendix~\ref{app: HOP parameter space}), and this should be kept in mind when interpreting our results to disentangle effects arising from the adopted methodology from those reflecting real physical trends (we defer a more detailed discussion of this point in Section~\ref{sec: what is a clump}).

In the present analysis, we proceed in two steps. First, we selected a subsample of Hi-GAL star-forming sources sharing a similar distance to limit the impact of physical resolution on their properties. So, we divided the Hi-GAL sample into heliocentric distance bins of $1\,\mathrm{kpc}$ each (the size of our domain). 
We then retained only those bins in which the fraction of structures with a diameter $\leq 0.3\,\pc$ ($\approx 10$ resolution elements for our simulations) was less than 5\%. This ensures that the comparison focuses on structures that, in the simulations, are sufficiently resolved to capture their internal physics.
The closest bin that satisfies this condition -- and guarantees the highest spatial resolution -- contains the clumps between $7$ and $8\,\mathrm{kpc}$, with a total of 1266 star-forming sources.

Once selected the Hi-GAL sources, we fine-tuned the \texttt{HOP} parameters so that the average sphericised radius of the structures\footnote{This is defined as $R\equiv\left(3V\,/\,4\pi\right)^{1/3}$, with $V$ the clump volume. This, to some extent, mimics the radius computed for Hi-GAL clumps, which is recovered similarly in terms of surface area.} resembles that of the observed subsample ($\approx0.38\,\pc$). 
After a parameter search, we set\footnote{As mentioned above, the choice of the parameter is not unique. Increasing (or lowering) the threshold $n_\mathrm{thr}$ can be compensated, to some extent, by imposing a lower (or higher) saddle density. Although the clump size distribution varied visibly,  their mass distribution and star-forming properties remained resilient to these changes. See also Appendix~\ref{app: HOP parameter space}.} $n_\mathrm{thr}=3\times10^3\,\ndens$, $n_\mathrm{saddle}=10\,n_\mathrm{thr}$ and $n_\mathrm{peak}=2\,n_\mathrm{thr}$ (using the relative parameter scaling also adopted in \citealt{Colman2024}). 
We note, however, that the choice of the $n_\mathrm{peak}$ parameter has little effect on the results as long as $n_\mathrm{peak}\ll n_\mathrm{sink}$. Indeed, we focused only on star-forming clumps -- that is, clumps containing at least one sink that has accreted throughout the last timestep. By construction, if a sink is accreting, the clump has reached the sink formation density $n_\mathrm{sink}$ \citep{Bleuler2014}.

\begin{figure}
    \centering
    \includegraphics[width=1\linewidth]{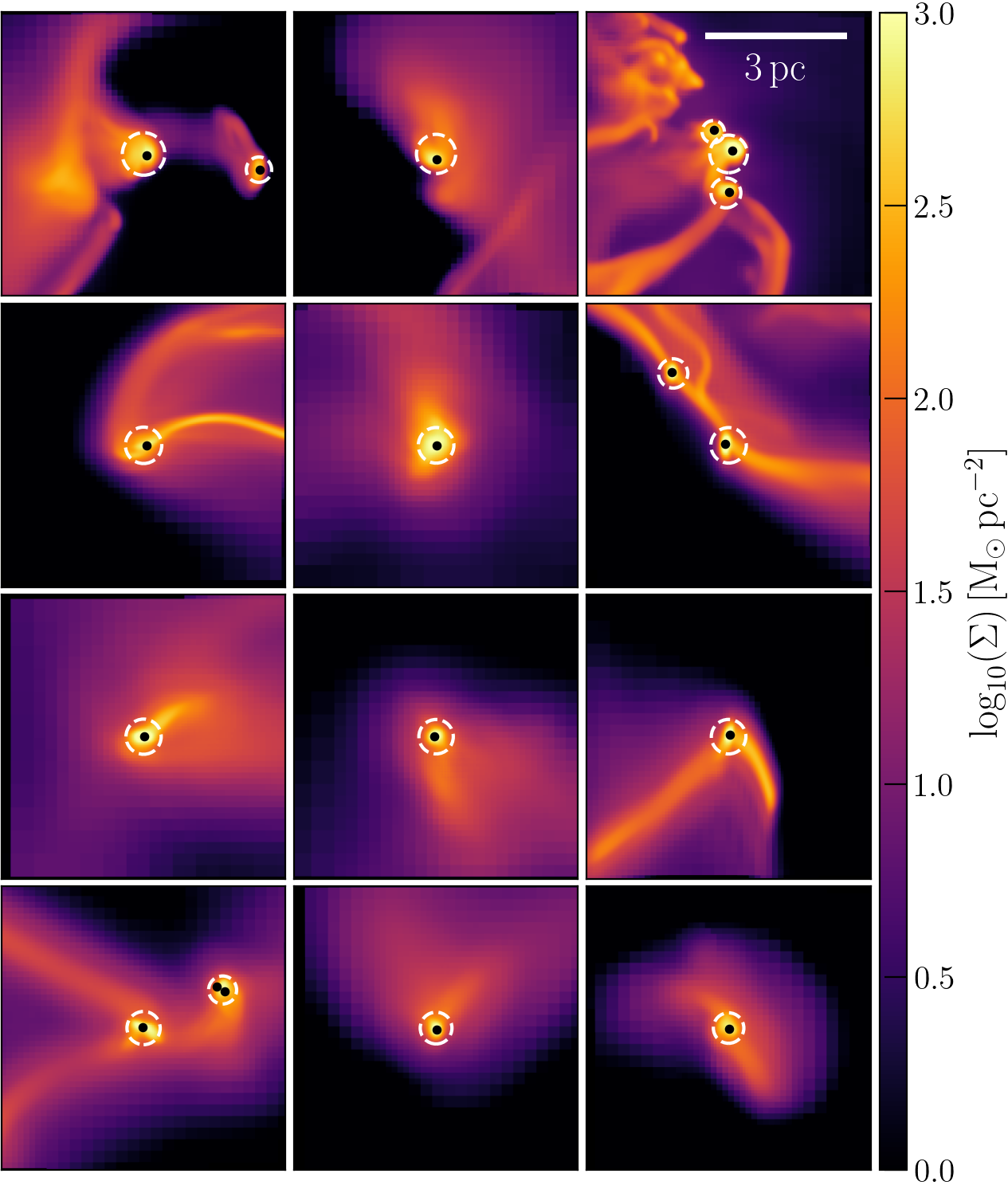}
    \caption{Collection of clouds developing in \refsim simulations, hosting one or multiple star-forming clumps (the black dots mark the accreting sink particles). The white circles highlight the sphericised radii of the recovered clumps. Each box covers a window $8\,\pc\,\times8\,\pc$.}
    \label{fig: set of clumps}
\end{figure}

In this work, we consider the onset of the SF as a reference point ($t_\star$). The selected snapshots are all taken $\approx4.5\,\Myr$ after $t_\star$, which is multiple times the free-fall time at the selected densities ($\tff\approx0.8\,\Myr$ at $3000\,\ndens$). By this time, the SFR in the box has already reached a steady state settling around $0.0004$, $0.0012$, $0.0049$, $0.0166$ and  $0.0174\,\MSun\,\mathrm{yr^{-1}}\,\mathrm{kpc^{-2}}$ for the \highturb, \outflowsim, \refsim, \lowturb and \highn runs respectively (see Section~\ref{sec: global star formation rate} for a more in-depth analysis and comparison with observations). Importantly, we verified that the clump size and mass distributions have also nearly reached a steady state in every simulation (see Appendix~\ref{app: temporal evolution}). We note that, in the current analysis, the runs themselves have already reached $\geq 165\,\Myr$, so we can be confident that the populations are not affected by the initial idealised setup. 
To give a visual idea of the structures extracted from the simulations, we show in Fig.~\ref{fig: set of clumps} a sample of star-forming clumps extracted from the \refsim simulation. 
With these parameters, the selected snapshots in \refsim, \lowturb, \highturb, \highn, and \outflowsim simulations contain, respectively, 69, 117, 7, 79 and 79 \texttt{HOP} star-forming clumps.

\begin{figure*}[t]
\includegraphics[width=1 \linewidth]{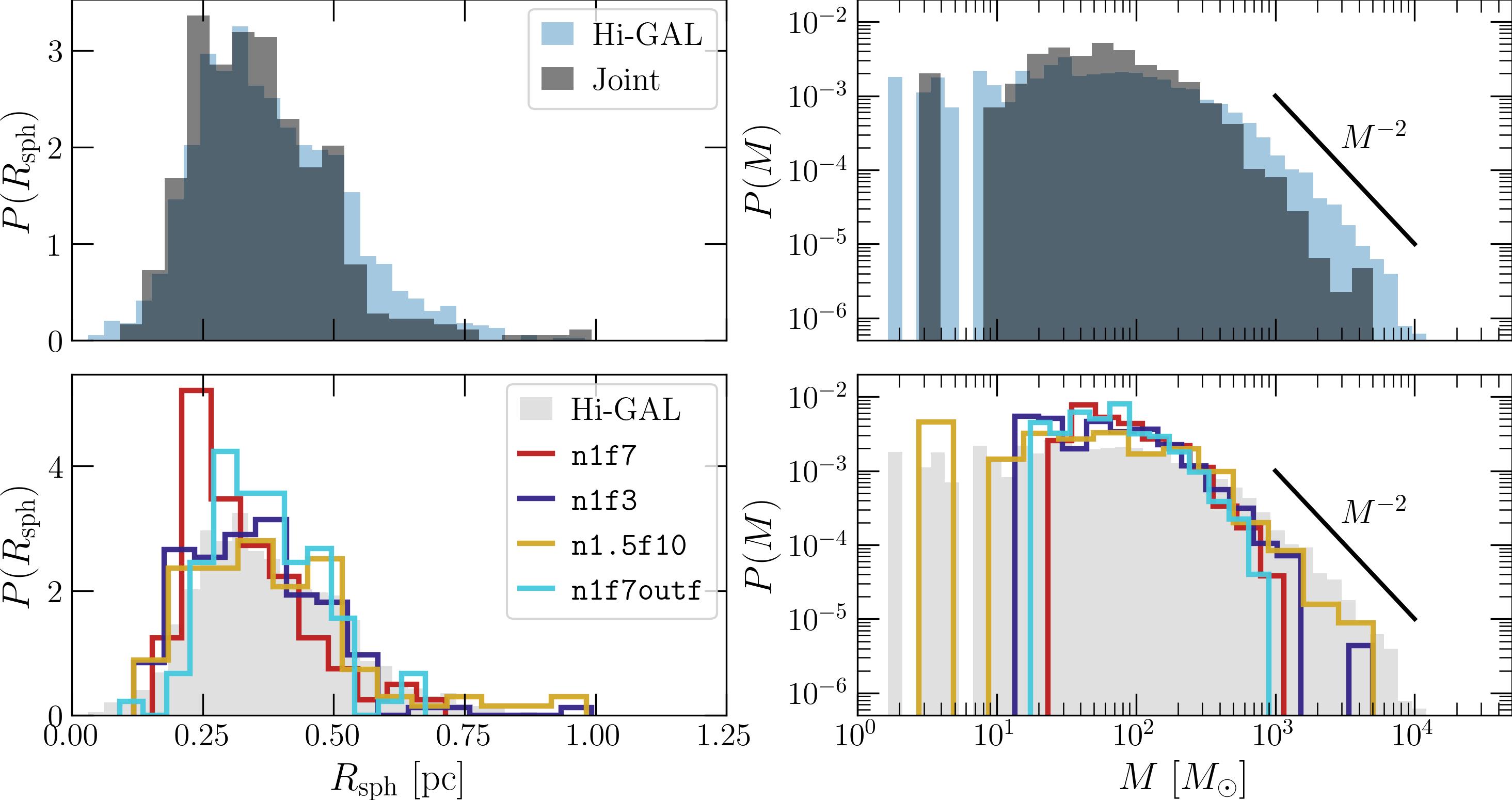}
\caption{Size (left column) and mass (right column) distributions from our simulated sample, compared to the Hi-GAL ones. The top row shows the stacked distribution of simulations without outflows (dark grey), which closely resembles that of the selected Hi-GAL subsample (light blue). The bottom row, instead, shows the distributions from each simulation separately (red: \refsim; dark blue: \lowturb; yellow: \highn; cyan: \outflowsim), along with the ones from Hi-GAL (grey). The mass and radius of the seven \highturb clumps (not shown due to the limited sample) range between $70$ and $320\,\MSun$, and $0.33$ and $0.46\,\pc$, with averages of $176\,\MSun$ and $0.38\,\pc$.}  
\label{fig: R and M distributions}
\end{figure*}
\section{Results}
\label{sec: results}

\subsection{Property distributions}
\label{sec: property distributions}

We begin by comparing the distributions of the \texttt{HOP} clump properties with those of the selected Hi-GAL sources. Figure~\ref{fig: R and M distributions} shows the joint radius (left) and mass (right) distributions obtained by stacking the non-jet simulations (top row), together with the corresponding distributions for each individual run (bottom row).
Both joint distributions of radius (top) and mass (bottom) resemble the Hi-GAL distributions, matching the ranges ($R\in[0.1,\,1]\,\pc$; $M\in[1,\,10^4]\,\MSun$), the visual shape, and the position of the peak. The simulated mass distribution declines more rapidly towards the high-mass end, but this is likely due to both the limited statistics and the narrower range of environments sampled by our simulation suite compared to the Galaxy-wide Hi-GAL survey.
We point out that, because the \texttt{HOP} parameters were calibrated to reproduce the median size of the Hi-GAL clumps, agreement near the centre of the $R_\mathrm{sph}$ distribution is expected, since the calibration effectively shifts the bulk of the distribution towards the observed characteristic scale.
However, matching this scale does not constrain the masses of individual structures or the detailed shape of the distribution (particularly its tails). These properties emerge from the local morphology and fragmentation of the density field\footnote{Consider, for example, a simulation whose incorrect fragmentation physics produces excessively dense clumps. Adjusting $n_\mathrm{thr}$ so that their characteristic sizes match those of the Hi-GAL clumps would still yield excessively massive structures, since, approximately, $M\propto R^3\langle n\rangle$. Thus, calibrating the characteristic size cannot by itself guarantee agreement in mass or in the full distribution.}.

Some differences are visible when comparing the separate simulated distributions. The \refsim run (red) shows a tendency to form smaller and less massive clumps than \lowturb (dark blue) and \highn (yellow). Conversely, the \highn run produces fewer clumps, which span larger ranges of radii and masses (from $\approx 1$ to $10^4\,\MSun$). In this plot, data from the \highturb run are not shown. The simulation contains only 7 clumps at the selected snapshot (though the number is roughly preserved at other snapshots). Nevertheless, this \highturb set remains in agreement with the other realisations, with its clumps' masses and radii spanning between $70$ and $320\,\MSun$, and $0.33$ and $0.46\,\pc$ (with averages of $176\,\MSun$ and $0.38\,\pc$). 

In any run without outflows, the peak and the ranges of the radius distribution remain in rough agreement with that observed in the Hi-GAL selected sample, and the mass distributions show a plateau at around $10^{2}\,\MSun$, which is close to that of the observed set. While \lowturb and \highn show a close match with the mass distribution of Hi-GAL clumps, \refsim clumps display less variation with a more limited range (this likely contributes to the steeper joint mass distribution). 
To compare the agreement between the distribution of the observed and simulated populations, we performed a maximum-likelihood fit on the high-mass tail of the distribution, $P(M)\sim M^{-a}$. For each set, we allowed both the lower and upper mass limits of the fitted interval to vary, adopting the values that minimised the Kolmogorov–Smirnov distance between the data and the model. With this method, we find similar logarithmic slopes in all the samples: $-2.2\pm0.1$ for the Hi-GAL sample, $-2.2\pm0.3$ in \lowturb, $-2.3\pm0.2$ in \highn, and $-2.7\pm0.3$ in \refsim (where the steepness is probably due to the cut-off at masses $\gtrsim10^3\,\MSun$).

Finally, we can observe the effects of outflows by comparing the distributions of \outflowsim clumps with those from the sibling \refsim. We note that outflows slightly decrease the clump mass by roughly by 15\%, decreasing the $\langle M\rangle$ from $\approx 200$ to $167\,\MSun$. More evidently, this feedback increases the average clump size, but also introduces a sharper cut-off at higher radii. These effects can be explained by the local dynamical action the flows have on the structures. By reinjecting energy into the medium, they also increase the turbulent support, leading to larger clumps. At the same time, the ejection may also increase fragmentation in the largest and objects (as also reported in \citealt{Federrath2014, Guszejnov2023}) and drive the expulsion of material, leading to a decrease in mass and a lack of large structures.

\subsection{Global star formation rate}
\label{sec: global star formation rate}
Before moving on to analyse the SF properties of single clumps, we performed a quick sanity check calculating the total SFR within the kpc boxes and comparing this with what is expected in Milky Way disc environments. \citet{Elia2025} provide a map of the surface SFR of the Milky Way exploiting the extensive coverage of the Hi-GAL sample. 
Adopting the semi-analytical formula tested in \citet{Elia2022}, they associate an SFR with each source as:
\begin{equation}
    \mathrm{SFR} = 0.56\,\left(\frac{M}{\MSun}\right)^{0.74\pm0.03}\,\MSun\,\Myr^{-1},
\label{eq: Hi-GAL SFR}
\end{equation}
with $M$ the clump's mass. This conversion is an operational definition of the method adopted by \citet{Veneziani2017}, in which the SFR was estimated using the \citet{Molinari2008} evolutionary tracks. Although this is not equivalent to a direct comparison with an observed SFR\footnote{In fact, \citet{Elia2025} showed how direct estimates recovered from the $70\,\mu$m emission drastically underestimate the SFR of the single sources, and only when the total emission over a broader area is considered the total Galactic SFR returns consistent with other measures.
}, 
\citet{Elia2022} showed that the Galactic SFR, estimated by summing Eq.~\ref{eq: Hi-GAL SFR} over the whole Hi-GAL sources catalogue, is in agreement with the Galactic SFR obtained through other observational means. 
According to their study, the Galactic SFR per unit density between $6$ and $8\,\mathrm{kpc}$ from the Galactic centre varies between 0.004 and $0.009\,\MSun\,\mathrm{yr}^{-1}\,\mathrm{kpc^{-2}}$. However, strong fluctuations are visible in the Galactic SFR surface density map at $0.25\,\mathrm{kpc^2}$ resolution (see their Fig.~5). To make a better comparison with the SFR of our kpc boxes, we resampled the map to a pixel size of $1\,\mathrm{kpc}$, finding that within the same annulus the SFR surface density can reach values as high as $0.025\,\MSun\,\mathrm{yr}^{-1}\,\mathrm{kpc^{-2}}$ and as low as $0.002\,\MSun\,\mathrm{yr}^{-1}\,\mathrm{kpc^{-2}}$.

In our simulations, we cannot trace the clumps' SFR directly, as the finite resolution prevents us from capturing the gas dynamics beyond $n_\mathrm{sink}$. As mentioned above, the sinks we produce are essentially dense-gas containers, composed of both stars and gas. The common (often implicit) assumption in numerical works is that once the gas reaches such densities, gravity will eventually lead to the material's total collapse \citep[see also][]{Bate1995, Federrath2010, Clark2014}. In this way, the mass converted into sinks per unit time, $\simuSFR$, serves as a proxy for the actual SFR. Throughout the rest of the paper, we denote the simulated SFR with $\simuSFR$ and reserve the term `SFR' for observational estimates or for discussions of the physical quantity itself.

For the present comparison with observations, we adopt $\mathrm{SFR}\equiv\simuSFR$.
For each \texttt{HOP} clump, we identify the sinks it contains and estimate $\simuSFR$ by summing the mass accreted onto those sinks between two consecutive outputs and then dividing it by the snapshots' temporal separation ($\Delta t\approx0.1\,\Myr$). Summing over the clumps $\simuSFR$, we obtain that the instantaneous total $\simuSFR$ in the domain at the selected snapshots is roughly $0.005$, $0.017$, $0.017$ and $0.0004\,\MSun\,\mathrm{yr}^{-1}\,\mathrm{kpc^{-2}}$ in \refsim, \lowturb, \highn, and \highturb, respectively\footnote{Although being instantaneous values, we note that after $1\,\Myr$ from the formation of the first sink, the systems stabilise. Relative fluctuations in the total $\simuSFR$ remain of the order of $\approx \pm50\%$. See also Fig.~5 in \citetalias{Hennebelle2026}.}. 
In \outflowsim, outflows reduce the overall $\simuSFR$ of \refsim by a factor of $\approx 4.2$. Therefore, the \highturb and \outflowsim simulations tend to underestimate the average SFR of the Milky Way, whereas \highn and \lowturb tend to overestimate it -- although in these latter cases the deviation remains within the range of fluctuations observed in the SFR surface density map of \citet{Elia2025}.

\subsection{Clump star formation rate}
\label{sec: star formation properties}

\begin{figure}
\includegraphics[width=1 \linewidth]{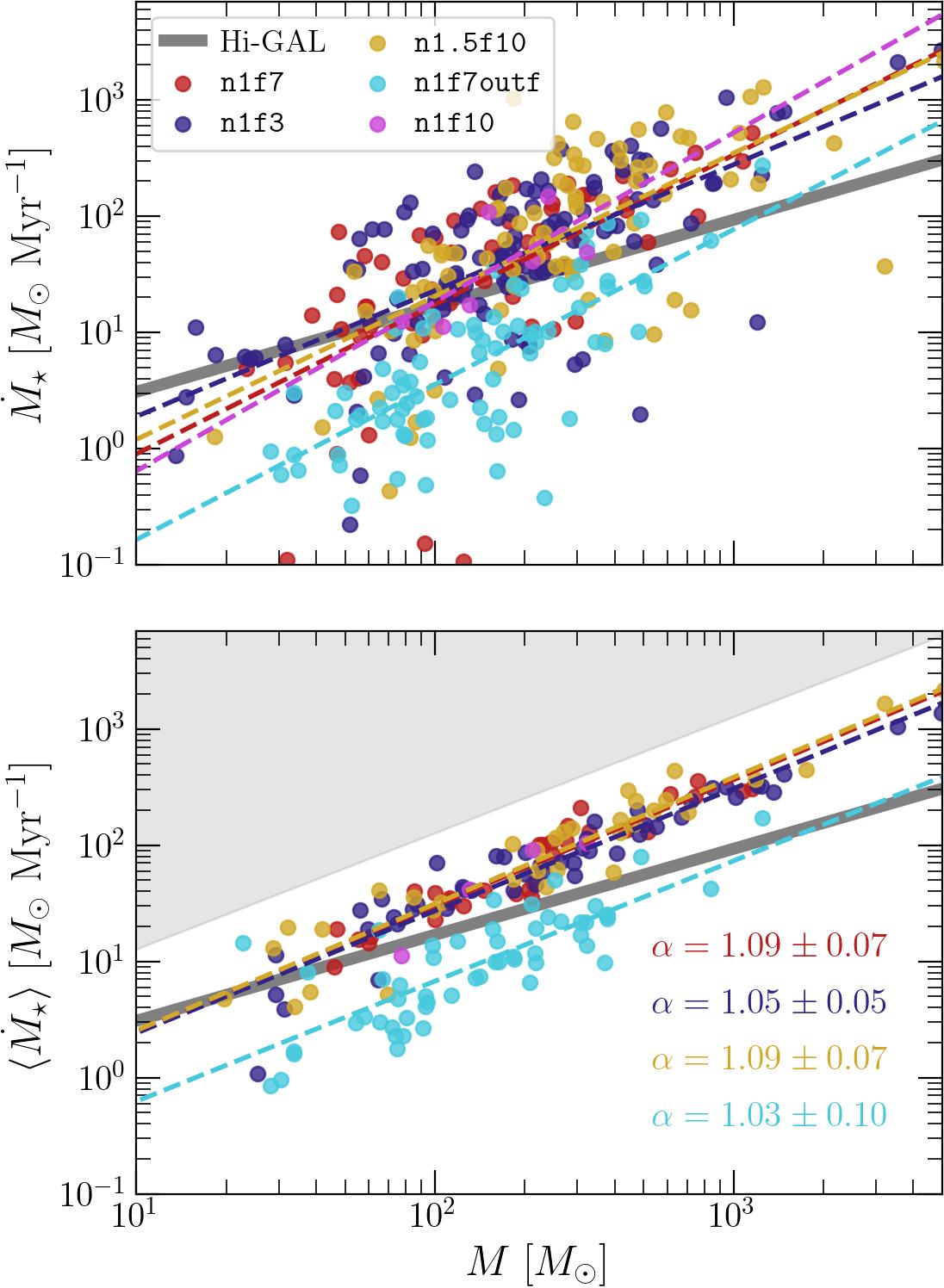}
\caption{Relationship between the clumps' $\simuSFR$ and their mass. The top graphs show the distribution for the whole sample at the selected snapshots (red: \refsim; dark blue: \lowturb; yellow: \highn; purple: \highturb; cyan: \outflowsim), with the grey line displaying the analytical estimates for Hi-GAL clumps (Eq.~\ref{eq: Hi-GAL SFR}). The dashed lines show the best power-law fit to the various populations. In the bottom panel, the average $\langle\simuSFR\rangle$ (estimated as the total enclosed sink mass divided by the age of the oldest sink) is used instead, with the sample reduced to only clumps older than one free-fall time ($\approx 0.8\,\Myr$). As the total clump mass includes that of the sinks (see Section~\ref{sec: methods}) and $\tau_\star\geq\tff$, the shaded area marks the theoretically forbidden region ($\langle\simuSFR\rangle>M/\tff$). The dashed lines report the best fit (see Table~\ref{tab:sfr_fits_alpha_only}).}
\label{fig: SFR vs M}
\end{figure}

As stated above, we can extract for each clump the internal $\simuSFR$ simply by looking at the accretion rates from enclosed sinks. However, one may note that the Hi-GAL SFR derived from Eq.~\ref{eq: Hi-GAL SFR} only give their SFR averaged over the theoretical lifetime of the clump rather than an instantaneous measure (see \citealt{Elia2022} for details). Therefore, to achieve a closer and more meaningful comparison, we need to average the clumps' $\simuSFR$ over their star-forming phase. We estimate this by dividing the total sink mass contained in a clump by the age of the oldest sink\footnote{All the clumps in the sample contain at least one actively accreting sink (Section~\ref{sec: methods}), so the star-forming duration inferred from the oldest enclosed sink is a lower limit. Furthermore, we point out that this strategy does not account for the clump's history, which may have accreted other structures along with its sinks over time. Aside from these details, for which one would need to track the evolution of these structures throughout the simulations, this method still provides a practical $\langle\simuSFR\rangle$ proxy on Myr timescales, comparable with the lifetime-average of Hi-GAL clump estimates.}, $\langle \simuSFR\rangle=M_\star/\tau_\star$.
Finally, we exclude clumps younger than one free-fall time ($\approx0.8\,\Myr$ at a density of $3\times10^3\,\ndens$), to remove possible fluctuations at the beginning of the collapse. 

Figure~\ref{fig: SFR vs M} shows the $M$--$\simuSFR$ relation obtained for \texttt{HOP} clumps in the various simulations at the reference snapshots. The top panel displays the distribution of the instantaneous $\simuSFR$ for the whole sample, while the bottom panel shows the subsample of older clumps, along with their averaged $\langle\simuSFR\rangle$. As is evident, the averaging procedure polishes the data, as the scatter observed in the top panel is in part due to fluctuations in $\simuSFR$ that clumps may experience over time. 
Comparing with Hi-GAL expected SFRs (grey line), we see that the bulk of distributions for simulations without outflows lie at slightly higher values of $\simuSFR$. Nonetheless, we recall that this line derives from semi-analytical models where the clumps are assumed to evolve in isolation and are not derived from direct measures. Therefore, the relation should be regarded as a general guideline for the bulk of the distribution. Protostellar outflows (cyan) decrease the average $\simuSFR$ by a factor of $4$ relative to the sibling \refsim run (red; cf. last column of Table~\ref{tab:sfr_fits_alpha_only}). This moves the simulated clumps in the direction of the Hi-GAL trend, but it is apparently too effective, yielding to clumps that are systematically less efficient than the observed ones.

\begin{table*}
\renewcommand{\arraystretch}{1.1} 
\caption{Sample sizes and best-fit parameters of the mass--SFR relation.}
\label{tab:sfr_fits_alpha_only}
\setlength{\tabcolsep}{8pt}
  \centering
\begin{tabular}{l c c c c c}
  \hline\hline
  Sample & $N$ & $N_\mathrm{old}$ & $\alpha$ (Instantaneous $\simuSFR$) & $\alpha$ ($M_\star/\tau_\star$) & $\langle \simuSFR\rangle\ \mathrm{at}\ 100\,\MSun\ [\MSun\,\Myr^{-1}]$ \\
  \hline\hline
  \noalign{\vskip 0.1cm}
    Hi-GAL        & 1266     & --- &       ---      & $0.74\pm0.03$ & $17$\\
    \refsim       &  69      &  30 &  $1.29\pm0.19$ & $1.09\pm0.07$ & $29$\\
    \lowturb      &  116     &  27 &  $1.15\pm0.15$ & $1.05\pm0.05$ & $31$\\
    \highn        &  79      &  36 &  $1.23\pm0.16$ & $1.09\pm0.07$ & $37$\\
    \outflowsim   &  7       &  49 &  $1.33\pm0.14$ & $1.03\pm0.10$ & $7$\\
    \highturb     &  79      &  54 &  $1.46\pm0.66$ & --- & $21^{*}$\\
    \hline\hline
  \end{tabular}
  \vspace{0.2cm}
  \tablefoot{First column: name of the sample. Second column: size of the selected subsample of star-forming clumps. Third column: number of star-forming clumps with the age of the oldest sink greater than one free-fall time ($0.8\,\Myr$). Fourth column: fitted slope using the clumps' instantaneous $\simuSFR$. Fifth column: slope obtained by averaging over the clump star-forming phase (total mass of enclosed sinks divided by the age of the oldest sink). For \highturb performing a reliable fit was not possible due to the scarcity of points (4). Sixth column: $\simuSFR$ at $100\,\MSun$, computed through the power-law fitted to the oldest sample .\\(*) Calculated imposing a linear relation. }
  \label{tab:sfr_fits_alpha_only}
\end{table*}

What catches the eye the most, however, is the firm consistency across the simulations without outflows, especially in the $\langle\simuSFR\rangle$ case. Fitting a power law relation on the distributions, we obtain that in both panels all the clumps distribute along the same relationship, with the slope $\alpha$ being compatible across the realisations (Table~\ref{tab:sfr_fits_alpha_only}). We also note that the slope decreases in the fit done with $\langle\simuSFR\rangle$ due to decreased vertical scatter (third column). The result is a nearly linear relation between the two clump variables, which is in very good agreement with what is reported for Galactic HCN clumps \citep{Liu2016}. This implies that, recalling what was observed in Section~\ref{sec: global star formation rate}, the large-scale properties of the medium must act on the general properties of the clump population (such as numerosity and mass and size distributions, see second column in Table~\ref{tab:sfr_fits_alpha_only}), but have little influence on the SF process at small scales. 
Additionally, we note that the presence of outflows does not affect the fitted slope, which remains close to that obtained for \refsim clumps.

\subsection{Star formation efficiency per free-fall time}
\label{sec: sfeff}
Once the $\simuSFR$ is obtained for each clump, a quantity that is easy to recover (and yet insightful) is their SF efficiency per free-fall time \citep{Krumholz2005}. This is defined as the fraction between the actual SFR measured and the one expected if the clump were free-falling:
\begin{equation}
    \eff\equiv\frac{\mathrm{SFR}}{\dot{M}_\mathrm{ff}}=\frac{\mathrm{SFR}}{M}\,\tff,
\label{eq: eff}
\end{equation}
with $\tff$ the clump free-fall time.
Using the uniform-sphere approximation, the free-fall time can be inferred for each Hi-GAL source, whereas for \texttt{HOP} clumps it suffices to use their average density.

\begin{figure}
\includegraphics[width=1 \linewidth]{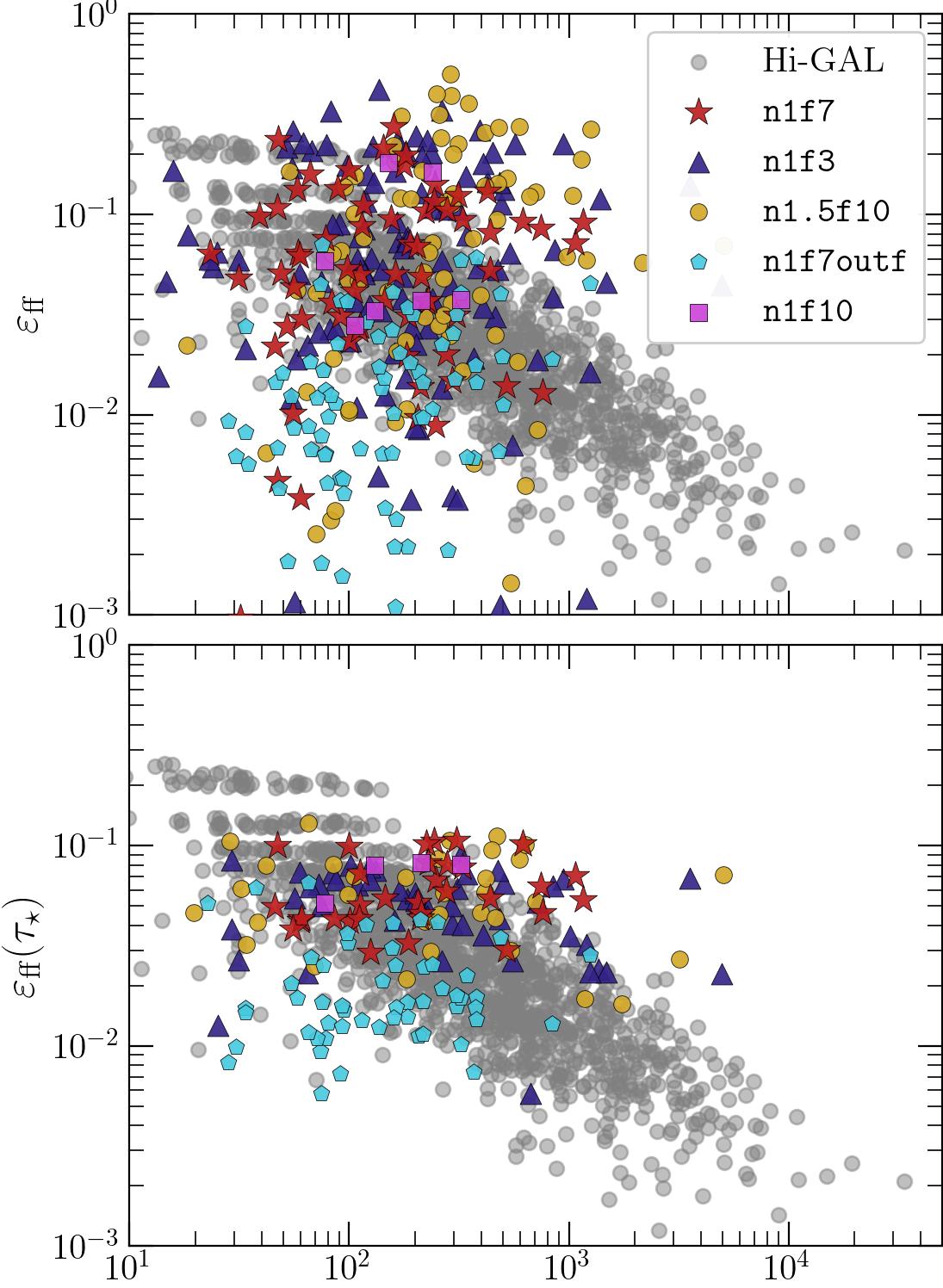}\\
\caption{Distribution of the clumps' $\eff$ as a function of their mass, for the simulated set (star: \refsim; triangle: \lowturb; circle: \highn; pentagon: \highturb; square: \outflowsim; the colour code is preserved from previous Figs.) and Hi-GAL sources (grey circles). In the top panel the instantaneous $\simuSFR$ is used, while in the bottom one the $\simuSFR$ is averaged over the oldest sink age (see the text and Fig.~\ref{fig: SFR vs M}).
}  
\label{fig: SFEff}
\end{figure}

Figure~\ref{fig: SFEff} displays the distribution of $\eff$ as a function of the clump mass. As in Fig.~\ref{fig: SFR vs M}, we plot in the top panel the clumps $\eff$ estimated using the instantaneous $\simuSFR$ in Eq.~\ref{eq: eff}, while in the bottom panel we adopt $\langle \simuSFR\rangle$ for the population of clumps older than one $\tff$ to compute the same parameter -- which we denote as $\eff(\tau_\star)$ for clarity. In this graph, Hi-GAL clumps follow on a negative slope of $\approx-3/4$ as expected from Eqs.~\ref{eq: Hi-GAL SFR}-\ref{eq: eff}: using the uniform sphere approximation, we can derive $\eff\sim M^{-0.76}\,R^{3/2}$. Since the radii barely span one order of magnitude, the dependence on $M(R)$ is blended into the dispersion of the points. 

Simulations without outflows tend to have higher $\eff$ than Hi-GAL clumps, with the median value shifted by a factor of $\gtrsim 2$. As visible in Table~\ref{tab:epsff_median}, the median value of the parameter does not vary significantly between its instantaneous and time-averaged formulation.
Similarly to the Fig.~\ref{fig: SFR vs M}, the scatter is large when $\simuSFR$ is employed, with $\eff$ spanning within $[10^{-3},0.5]$ across simulations. The scatter is greatly reduced in $\eff(\tau_\star)$, with the $1\sigma$ dispersion being $0.16\,\mathrm{dex}$ in the \refsim simulation, and $0.23\,\mathrm{dex}$ in all the others (without considering \highturb). Furthermore, the populations display little to no dependence of the parameter on the mass. Fitting a power law on $\eff(\tau_\star)\sim M^\eta$ returns slopes $\eta\approx 0.08\pm0.08$ in \refsim and $-0.08\pm0.09$ in both \lowturb and \highn.

Protostellar outflows clearly reduce the $\eff$ of simulated clumps, with the median value decreasing by a factor of 4.25 with respect to \refsim (similarly to what was noted in the previous Section). Despite the better match in the median value of $\eff$ (Table~\ref{tab:epsff_median}), we see that the clumps from \outflowsim significantly depart from the region covered by Hi-GAL clumps on the $(\log M,\log \eff)$ plane (particularly in the top panel), which may indicate that outflow feedback is too strong in our subgrid prescription.

\begin{table}
\renewcommand{\arraystretch}{1.7}
\setlength{\tabcolsep}{14pt}
\caption{Median $\eff$ in the analysed clump populations.}
  \centering
  \begin{tabular}{l c c}
    \hline\hline
    Sample  & $\varepsilon_{\rm ff,\,med}\times10^{-2}$ &
    $\varepsilon_{\rm ff,\,med}(\tau_\star)\times10^{-2}$\\
    \hline\hline
    Hi-GAL    & --- & $2.6^{+0.1}_{-0.1}$ \\
    \refsim   & $5.1^{+1.3}_{-0.3}$ & $5.3^{+0.6}_{-0.4}$ \\
    \lowturb  & $5.3^{+0.9}_{-0.4}$ & $5.4^{+0.5}_{-0.3}$ \\
    \highn    & $6.6^{+1.1}_{-0.7}$ & $6.0^{+1.1}_{-0.3}$ \\
    \highturb  & $3.7^{+2.1}_{-0.1}$ & $8^*$ \\
    \outflowsim   & $1.2^{+0.1}_{-0.2}$ & $1.6^{+0.1}_{-0.1}$ \\
    \hline\hline
  \end{tabular}
  \vspace{0.2cm}
  \tablefoot{First column: name of the sample. Second column: Median $\eff$ of the corresponding sample. Third column: fraction of simulated clumps lying within the Hi-GAL highest-density region in the $(\log M,\log\eff)$ plane (see the text). Uncertainties correspond to the $1\sigma$ bootstrap confidence interval.\\
  ($^*$) The median value is left for reference, but it should be taken with care given the few number of points (4).
  }
  \label{tab:epsff_median}
\end{table}

\section{Discussion}
\label{sec: discussion}
\subsection{What is a clump}
\label{sec: what is a clump}
One important point to discuss before putting the results into perspective is what type of structures we are looking at. From both observational and numerical perspectives, these clumps are recognised as overdensities relative to the mean field. 
When considering the full Hi-GAL sample, it is difficult to define a clear net transition between classes of structures. For example, the distinction between cores, clumps, and clouds proposed by \citet{Elia2021} relies solely on the structures' diameters, which tightly depends on the source heliocentric distance (due to physical resolution, cf. Fig.~4 in \citealt{Elia2017}). Similarly, in the set of simulations, the choice of the \texttt{HOP} parameters (mainly $n_\mathrm{thr}$ and $n_\mathrm{saddle}$) was not univocal (see Appendix~\ref{app: HOP parameter space}). 

\begin{figure}
    \centering
    \includegraphics[width=1\linewidth]{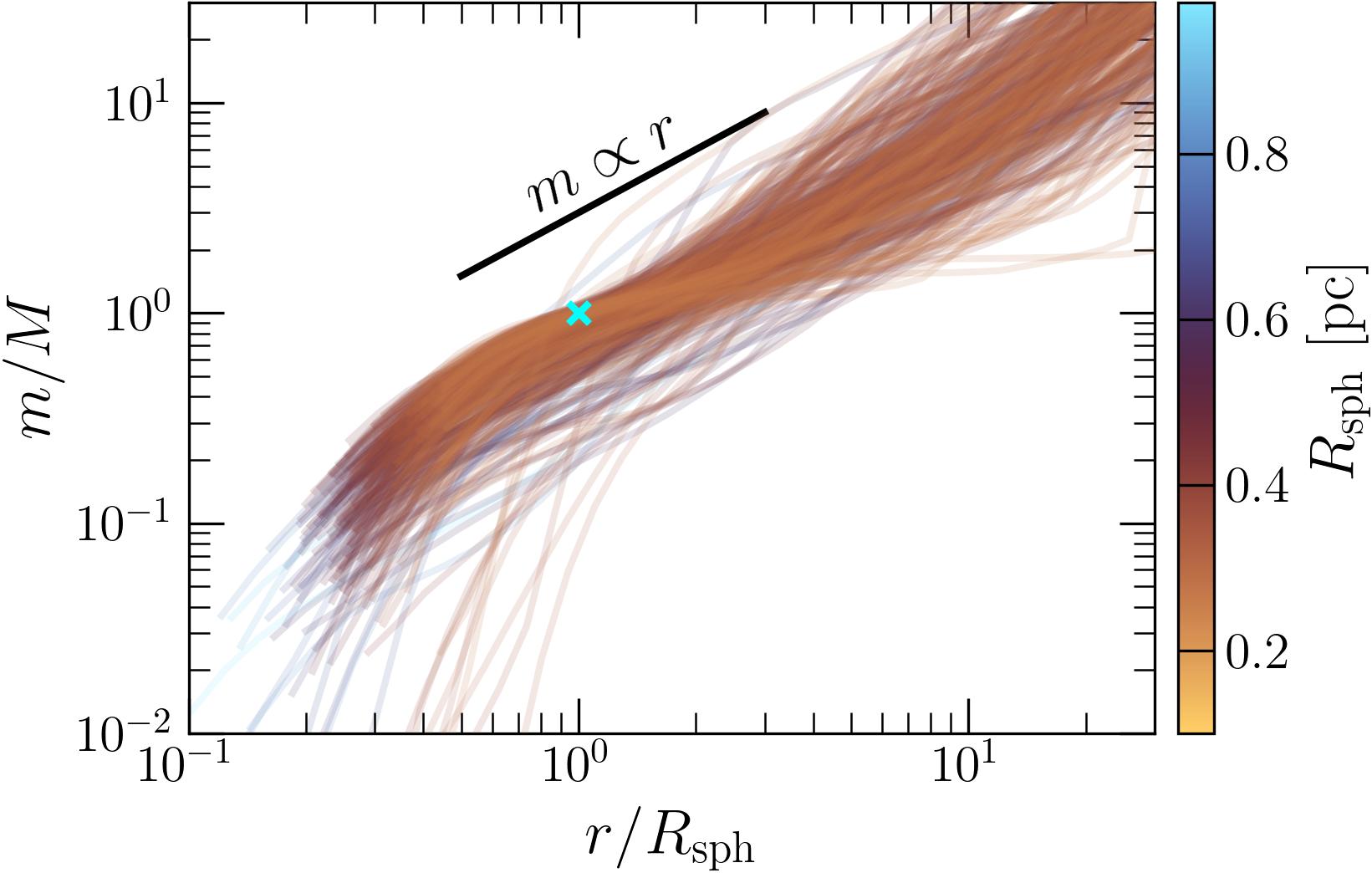}
    \\
    \includegraphics[width=1\linewidth]{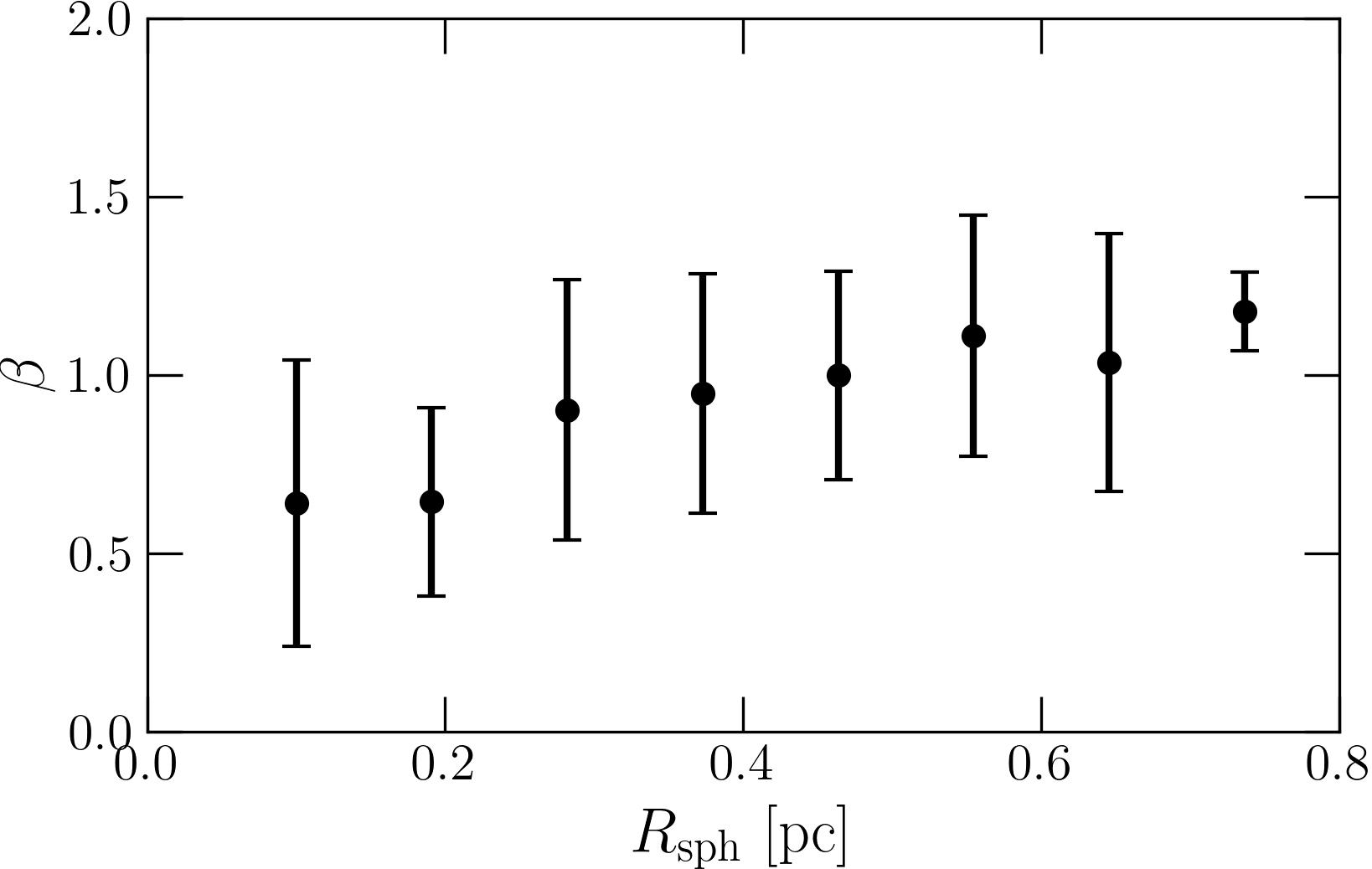}
    \caption{\textit{Top}: radial profile of the mass enclosed in concentric spheres of radius $r$ and centred in the simulated clumps' peak. Radii and masses are normalised by each clump’s spherical radius $R_\mathrm{sph}$ and mass $m$ (the cross point $r=R_\mathrm{sph}$, $m=M$). The mass accounts for possible sink particles within the spheres. 
    \textit{Bottom}: average logarithmic slope of the mass radial profile estimated between $r=R_\mathrm{sph}$ and $r=5R_\mathrm{sph}$, binned by the clump size. Error bars show the standard deviation in each $R_\mathrm{sph}$ bin.}
    \label{fig: radial profile}
\end{figure}

To better discuss this point, we plot in Fig.~\ref{fig: radial profile} (top panel) the radial profile of the mass around the peak of the \texttt{HOP} clumps, scaled with the clumps' sphericised radius $R_\mathrm{sph}$ and their \texttt{HOP} mass. While the steep profile at radii smaller than $R_\mathrm{sph}$ likely comes from artifacts due to the numerical resolution, we see that the mass steadily increases as we move away from the density peak. Fitting the profile's logarithmic slope as $m\sim r^{\beta}$ (bottom panel), we find that the mass growth resembles that expected for an isothermal sphere ($m\propto r$), as reported in multiple observational studies \citep[cf., for example,][]{Mueller2002, Butler2012, Hoang2025}. Numerical effects start to play a role\footnote{Stellar feedback also contributes to this effect. Once an \hii region impacts a clump, it erodes the outer layers, peeling off the outer, lowly-bounded material and imposing a steeper profile (see the top right panel of Fig.~\ref{fig: set of clumps}). Although we observe a general decrease in the clump radius in clumps embedded by ionised bubbles, we verify that only $10\%$ of clumps with $R_\mathrm{sph}<0.2\,\pc$ are affected by external radiation, supporting the conclusion that most of the observed flattening is indeed caused by numerical effects.} below $R_\mathrm{sph}\approx 0.25\,\pc$, when the slope $\beta$ starts to decrease: clumps with low $R_\mathrm{sph}$ display a net transition into the background around $R_\mathrm{sph}$. In any case, this sub-group of small clumps only contributes to $7\%$ of the total sample of \texttt{HOP} clumps (cf.~Fig.~\ref{fig: R and M distributions}) and therefore does not affect the validity of the conclusions discussed below.

Since the mass increases smoothly with radius, we cannot detect a sharp transition from a ``structure'' to the ambient medium in clumps resolved by the simulation ($R_\mathrm{sph}\gtrsim0.25\,\pc$).
In practice, the definition of `clump', even in simulations, remains somewhat arbitrary (in this case, defined merely by $n_\mathrm{thr}$) and indicates more to a regime of the SF process than distinct entities. We may therefore wonder whether analysing the SF laws of such ill-defined structures remains meaningful. However, several clues suggest that this is the case. The $M$--$\simuSFR$ relation exhibits a small intrinsic scatter (Sect.~\ref{fig: SFR vs M}), and it remains stable even when varying the \texttt{HOP} parameters  (Appendix~\ref{app: HOP parameter space}), showing that the scaling is not an artefact of our extraction method.

With the same reasoning, we can also understand the good agreement in the mass and size distributions. The \texttt{HOP} clumps identified in three dimensions and the Hi-GAL sources extracted in projection are not expected to correspond to identical physical entities. Nevertheless, once we match their characteristic spatial scale, the samples converge to similar mass-function slopes and a comparable plateau around $\sim10^2\,\MSun$ (cf. Fig.~\ref{fig: R and M distributions}), and they show a close visual agreement in the $R_{\rm sph}$ distribution.
Therefore, we can conclude that both procedures can be viewed as different ways of sampling the same underlying hierarchical density field produced by a gravoturbulent ISM.
In this sense, the difficult question of defining what a clump is no longer requires a solution. The measured scaling relations do not arise from well-defined physical objects; rather, they reflect the underlying self-similar structure of the star-forming medium (see also Appendix~\ref{app: HOP thresholds}).  

\subsection{Impact of the large-scale properties}
Comparing the simulations, we found that the large-scale environmental conditions play an important role in determining the properties of the resulting clump population. Because the support against gravitational fragmentation is smaller, the \lowturb simulation produces more star-forming clumps (cf.~Table~\ref{tab:sfr_fits_alpha_only}). For the opposite reason, in \highturb only a few clumps reach the critical density required to collapse and form sinks, as large-scale motions likely disperse the structures. Increasing the density balances the situation, with a population in \highn comparable in number to that of \refsim. 

Even in the mass and size distributions, we report some variations across the set. Although the distributions are visually similar, we see, for example, that \refsim clumps are, on average, more compact (cf. the size distribution in the lower-left panel of Fig.~\ref{fig: R and M distributions}). The mass distribution varies with the initial condition, mostly in the range covered by the population. Overall, the distributions from \lowturb and \highn, show closer agreement with the distributions observed in Hi-GAL clumps (Section~\ref{sec: property distributions}). Still, we notice that the high-mass tail preserves a similar slope between the simulations (which is also similar to that of Hi-GAL clumps), possibly hinting at a turbulent origin of the phenomenon\footnote{This is somewhat similar to what is observed for molecular clouds in the set of simulations analysed in \citet{Colman2024}.}.

These differences in clump populations translate into significantly different global $\simuSFR$ within the domain (Section~\ref{sec: global star formation rate}). The large population of \lowturb and the high efficiency of \highn clumps increases the total $\simuSFR$ by a factor of three compared to \refsim, while the turbulent support in \highturb significantly damps it.
Yet, the \texttt{HOP} structures follow a remarkably similar $M-\simuSFR$ relation across runs, meaning the large-scale properties mostly influence the forming clump population without affecting the SF process at small scales. In some sense, this justifies cloud-scale simulations, where the higher resolution allows a deeper analysis of the origin of this almost linear dependence. 

This weak dependence of the clump-scale SF law on large-scale conditions may seem at odds with recent observational studies reporting variations in this relation \citep[e.g.][]{Gallagher2018, Neumann2025}. A direct comparison, however, is not straightforward. Using HCN-based measurements in external galaxies, these works probe a much broader range of environments (from the dense and turbulent galactic centre to the more quiescent and sparse outskirts), whereas in our suite the mean density and turbulent velocity dispersion vary by less than a factor of two.
Possibly, probing more extreme environments in future works could provide insightful results. 
Furthermore, the physical resolution may also play an important role in these relationships. Averaging the HCN luminosity over the telescope beam, both star-forming and quiescent dense structures contribute. This may also affect the slope of the $M-\simuSFR$ relationship. This highlights the importance of high-resolution surveys for probing the SF process at small scales where individual clumps are more clearly separated.

\subsection{Clumps' star formation rate}
\label{sec: models on SFR}
We have seen that the SF process within the clumps behaves similarly across the simulations. In runs without outflows, all the \texttt{HOP} structures settle on the same power-law slope (Section~\ref{sec: star formation properties}). Even more importantly, the offset remains nearly identical across the four runs. As discussed above, this shows that once a given spatial scale is reached, the system becomes progressively insensitive to its large-scale environment and follows more universal physical scaling laws. 
Additionally, the almost-linear dependence of $\simuSFR$ on the clump mass, recovered by averaging over the star-forming phase of older clumps, is somewhat close to the operational exponent provided in Eq.~\ref{eq: Hi-GAL SFR} \citep[from][]{Elia2022}, and agrees with several observational studies that also report a linear relation between mass and SFR in dense structures \citep[e.g.,][and references therein]{Heiderman2010, Lada2010, Evans2014, Liu2016}, and with cloud-scale numerical simulations \citep{Priestley2026}.

We can construct a simple toy model to try deriving these scaling relations. We start by assuming that, for the long-lived clumps retained in our time-averaged sample, the sink accretion rate is of the same order as, or at least proportional to, the net inflow feeding the dense structure, $\simuSFR\sim\dot{M}_\mathrm{clump}$. This is a reasonable approximation for two reasons. First, the clumps we analysed persisted over several free-fall times, so their SF is not a brief burst associated with a single collapse event. Second, once we average over the star-forming phase, the $M-\simuSFR$ relation exhibits a relatively small scatter, suggesting that inflow and conversion into sinks remain approximately balanced on Myr timescales (Fig.~\ref{fig: SFR vs M}). For simplicity, in the following we adopt $\simuSFR=\dot{M}_\mathrm{clump}$, and we can write:
\begin{equation}
    \simuSFR\sim\dot{M}_\mathrm{clump}\sim C \rho_\mathrm{thr}R^2\,v_\mathrm{r},
    \label{eq: steady state}
\end{equation}
where $\rho_\mathrm{thr}=1.4\,m_\mathrm{p}\times n_\mathrm{thr}$ is the density at the interface defining the structure, $v_\mathrm{r}$ the infall velocity of the gas at the clump boundaries, and $C$ a constant that depends on the geometry of the system ($C=4\pi$ in a sphere).
With $R=(3M/4\pi\langle{\rho}\rangle)^{1/3}$, the only parameter left to set is $v_\mathrm{r}$.
For the infall speed at the boundaries, we consider two regimes. 

In a gravity-driven picture, the inflow is set by the clump potential well, so that $v_\mathrm{r}= f\,(GM/R)^{1/2}$, with $G$ the gravitational constant and $f$ a correction coefficient. 
In a completely free-falling scenario, $f=\sqrt{2}$. However, in a more realistic picture, part of the infall energy is redistributed into the other dimensions and converted into chaotic motion. This collapse can sustain a quasi-virial equilibrium \citep[e.g.][]{Hennebelle2012, Lee2016}, in which the energy injected by the infall and dissipated by the turbulence balance each other. For a virialised uniform sphere, the one-dimensional velocity dispersion is $\sigma_\mathrm{1D}=(GM/5R)^{1/2}$, which we use as a characteristic velocity-scale for $v_\mathrm{r}$ (imposing $f=\sqrt{1/5}$).

In a turbulence-dominated scenario, the inflow is instead regulated by the ambient velocity dispersion at the clump scale, for which $v_\mathrm{r}= \sigma_0(R/1\,\pc)^p$, with $\sigma_0$ the scaling parameter of the turbulence (defined from the large scales) and $p=1/3$ or $1/2$ depending on whether we consider sub or supersonic turbulence. In the following derivation, we adopt, for simplicity, the normalisation at $R=1\,\pc$ to $\sigma_0=1\,\kms$. This choice is close to the observed scaling relation: for example, \citet{Larson1981} reports $\sigma_0\simeq1.1\,\kms$, \citet{Solomon1987} a value of $\sigma_0\simeq0.76\,\kms$, and \citet{Heyer2004} $\sigma_0\simeq0.96\,\kms$.

With little algebra (see Appendix~\ref{app: derivation}), the equation above becomes:
\begin{equation}
\begin{aligned}
&\simuSFR(\mathrm{gravity}) \simeq
59\,\MSun\,\Myr^{-1}\,
\left(\frac{\gamma}{10}\right)^{-\frac{1}{2}}
\left(\frac{M}{100\,\MSun}\right)
\left(\frac{n_\mathrm{thr}}{3000\,\ndens}\right)^{\frac{1}{2}}
\\[4pt]
&\simuSFR(\mathrm{subsonic}) \simeq
71\,\MSun\,\Myr^{-1}\,
\left(\frac{\sigma_0}{1\,\kms}\right)\,
\left(\frac{\gamma}{10}\right)^{-\frac{7}{9}}
\\
&\hspace{4.58cm}\times
\left(\frac{M}{100\,\MSun}\right)^{\frac{7}{9}}\,
\left(\frac{n_\mathrm{thr}}{3000\,\ndens}\right)^{\frac{2}{9}}
\\[4pt]
&\simuSFR(\mathrm{supersonic}) \simeq
57\,\MSun\,\Myr^{-1}\,
\left(\frac{\sigma_0}{1\,\kms}\right)\,
\left(\frac{\gamma}{10}\right)^{-\frac{5}{6}}\\
&\hspace{4.58cm}\times
\left(\frac{M}{100\,\MSun}\right)^{\frac{5}{6}}\,
\left(\frac{n_\mathrm{thr}}{3000\,\ndens}\right)^{\frac{1}{6}}
\end{aligned}
\label{eq: accretion models with coeff}
\end{equation}
where we imposed $C=4\pi$ and $f=\sqrt{1/5}$, and defined the ratio between the average clump density and the \text{HOP} density threshold as $\gamma=\langle\rho\rangle/\rho_\mathrm{thr}$. Note that for a singular isothermal sphere $\gamma=3$ at all radii, but in the recovered \texttt{HOP} clumps the median of the distribution of this parameter settles around 10 in all simulations ($12.45$ in \refsim, 12.1 in \lowturb, 11.0 in \highn and 10.0 in \highturb).

In all cases, we obtain a slope close to unity, with a perfect match for gravitationally driven accretion. The main difference between the models is their dependence on $\rho_\mathrm{thr}$ and $\gamma$. We also note that the gravity-driven case effectively recovers the constant $\eff$ scenario ($\simuSFR=\eff M/\tff\sim \eff M\rho^{1/2}$), although preserving a different physical origin \citet{Lee2016}.

Comparing the normalisations with Fig.~\ref{fig: SFR vs M}, we see that these values are close to the observed $\simuSFR$ of \texttt{HOP} clumps. For $M=100\,\MSun$ the fit performed on the older population of clumps returns $\simuSFR\approx30-40\,\MSun\,\Myr^{-1}$ (cf. Table~\ref{tab:sfr_fits_alpha_only}). For our fiducial parameter choices, all three toy models predict values consistent with this estimate within a factor\footnote{This could imply $\dot{M}_\star\approx0.5\,\dot{M}_\mathrm{clump}$, with the possibility for the gaseous clump reservoir to grow over time.} of $\sim2-3$. 
However, the strictly linear slope observed in the $M-\simuSFR$ (Fig.~\ref{fig: SFR vs M}) relation and the relatively small dispersion observed in $\eff(\tau_\star)$ (Fig.~\ref{fig: SFEff}) seem to favour the gravity-driven picture.

Furthermore, in Appendix~\ref{app: HOP thresholds} we repeated the analysis varying the threshold $n_\mathrm{thr}$. The $M-\simuSFR$ relation remains consistently linear even adopting $n_\mathrm{thr}$ as low as $500\,\ndens$, indicating that the slope is not an artefact of the particular threshold choice. 
Interestingly, at the lowest threshold, $n_\mathrm{thr}=100\,\ndens$, the slope appears to flatten, possibly indicating a transition towards turbulence-regulated accretion, consistent with the findings of \citet{Brucy2025}.

\subsection{Clump $\eff$}
In all simulations, we observe a slope $\eff(\tau_\star)\sim M^\eta$ compatible with zero. This is expected from the gravity-regulated picture suggested in the previous Section, but turbulent accretion also predicts small negative slopes, with $\eta=-2/7\approx-0.29$ and $\eta=-1/6\approx-0.17$ in the subsonic and supersonic regimes, respectively. While this slope may rule out the subsonic turbulent model, only in \refsim do we have marginal evidence that gravity may dominate the accretion (as $\eta=0.08\pm0.09$). In \highn and \lowturb the fitted slope remains compatible with both scenarios  ($\eta=-0.08\pm0.09$). We also note that setting $\tff=\tff(\gamma\rho_\mathrm{thr})$ in Eq.~\ref{eq: eff}, we recover a predicted $\eff$ of roughly 0.14 at $M=100\,\MSun$ (independent of mass in the gravity-dominated scenario), which is within a factor of $2-3$ from \texttt{HOP} clump $\eff$ (considering the simulations without outflows). 

However, typical values reported on larger-scale studies are $\eff \sim 1-2\%$ (from kpc scales down to molecular-cloud scales; e.g. \citealt[][]{Krumholz2005, Pokhrel2021}). This is a factor of $5-10$ lower than the prediction, and a factor of 3 lower than the measured median of \texttt{HOP} clumps. This discrepancy may be reduced if we account for observational biases, such as telescope sensitivity and resolution, that may artificially lower the observed $\eff$ \citep{Suin2024}. The new JWST facility may represent a breakthrough to quantify this effect, possibly discovering a large number of previously undetected young stellar objects in highly embedded regions thanks to its extreme sensitivity \citep{Crompvoets2024, Lenkic2024}.
Additionally, this gap may be further reduced considering the effect of outflows. Indeed, Table~\ref{tab:epsff_median} shows that \outflowsim clumps can reach efficiencies even lower than what is expected from Hi-GAL clumps. Therefore, this disagreement may be fully explained by stellar feedback at clump scales.

Another interesting quantity to discuss is the scatter of $\eff$ among the same population. When $\eff$ is estimated using the instantaneous $\simuSFR$, the samples span more than two orders of magnitude (cf. Fig.~\ref{fig: SFEff}), but after averaging in time the dispersion drops greatly. This indicates that on small temporal scales ($\ll\tff$), the SF process is highly variable and subject to strong fluctuations (in agreement with \citealt{Grudic2019}). We also note that the median $\eff$ and its time-averaged version are close, probably because the simulated populations comprise several clumps at different evolutionary stages.

Observational measures are in part safeguarded from these fluctuations, as they cannot access the information regarding the instantaneous $\simuSFR$. Estimates of the local SFR are obtained through averages across a given temporal window -- e.g., $0.5\,\Myr$ when computing the SFR via counting of Class~0/I young stellar objects \citep[e.g.][]{Pokhrel2021}. In this sense, our results provide a natural explanation for the generally low scatter reported in observations. 
Indeed, it is promising to note that the scatter observed in the simulated samples (order of $0.16-0.23\,\mathrm{dex}$) is of the same order as the one observed in the local disc molecular clouds (see for example \citealt{Pokhrel2021} and \citealt{Hu2022}, where derivation of $\eff$ through young stellar object counting yielded to a dispersion $\approx 0.15\,\mathrm{dex}$)\footnote{Larger cloud-to-cloud dispersions have been reported for Galactic giant molecular cloud samples \citep[e.g.][]{Lee2016}. However, \citet{Pokhrel2021} argue that part of this difference may arise from systematics in SFR estimates based on massive-star tracers, such as free-free emission. Measurements in the Galactic Centre also show a larger scatter in $\eff$ \citep[e.g.][]{Barnes2017}, but we note that these regions probe physical conditions substantially different from those of the local Galactic disc considered here.}.

\subsection{The impact of protostellar outflows}
Regarding the impact of protostellar outflows, we found that the main effect is a reduction in $\simuSFR$ and $\eff$ by roughly a factor of 4.25. This value exceeds the loss from mass loading alone (a factor of 1.5), implying they dynamically act on the surroundings to reduce the local SFR (possibly through re-injecting turbulence). 
We note that this suppression of the SFR is stronger than what is reported in previous studies (generally around a factor of 2; \citealt{Nakamura2005, Federrath2015, Murray2018, Verliat2022, Appel2023}). This difference may partly arise from the lower resolution of the present simulations ($0.03,\pc$), compared to the simulations listed above, which generally reach resolutions about an order of magnitude higher. This may affect how the momentum injected by protostellar outflows is redistributed across the resolved scales, and therefore the quantitative strength of their impact. Nevertheless, the qualitative effects of outflows remain clearly identifiable in the mass and size distributions of the simulated clumps. In particular, we find that protostellar outflows tend to increase the average clump size, likely by enhancing the local turbulent support, while at the same time reducing their mass and limiting their maximum extension (Section~\ref{sec: property distributions}). This suggests that outflows promote additional fragmentation on local scales, as also reported in previous studies \citep{Federrath2014, Guszejnov2023}. 

Interestingly, we also found that outflows have little effect on the slope of the $M-\simuSFR$ relation, meaning they can slow the SF process, but have little effect on the mechanism that guides the collapse. This is also confirmed by previous studies conducted at cloud scales \citep{Suin2024}, which found that this mechanism does not affect the slope of the Kennicutt--Schmidt relation. 
Furthermore, the simulation with outflows shows a lower $\eff$ than Hi-GAL sources, with perhaps less overlap with the observed sample on the $(\log M-\log \eff)$ plane than in the other simulations. This could suggest an overrepresentation of the outflows, perhaps due to the selected mass-loading factor injected at the resolution limit of the simulations. In the current implementation, sinks eject at high speed ($\gtrsim 30\,\kms$) 1/3 of the mass they accrete prior to gas expulsion \citep[see][]{Verliat2022}. Although this estimate remains roughly consistent with observations, which place the proposed mass-loading range between 10 and 30\% \citep[][and references therein]{Frank2014}, a lower mass-loading fraction may better match the Hi-GAL $\eff(M)$ distribution. Nevertheless, regardless of the accuracy of this subgrid model, we can treat this effect as an upper limit on the outflows' impact, constraining future numerical exploration. 

\section{Conclusions}
\label{sec: conclusions}
We analysed the properties of star-forming structures in a set of simulated kpc-sized boxes and compared them with observed Hi-GAL clumps. After fine-tuning the clump-finding algorithm's parameters to match the structures' average size, we find good agreement in the mass and radius distributions. Together with the fact that the simulations reproduce the SF behaviour across scales (see Fig.~7 and Section~5 in \citetalias{Hennebelle2026}), this suggests that our numerical setup provides a reliable multi-scale framework to follow SF process from the kpc environment down to clump scales.
Exploiting the multi-scale coverage provided by the simulations, we found that:
\begin{itemize}
    \item The large-scale environment controls the clump population properties, but not the SF law within the clumps.    Indeed, the details of the mass and size distributions vary across simulations in peak position and range covered, as do the number of members. Given the tight relationship between clumps' $\simuSFR$ and mass, the box's global SFR is sensitive to the large-scale properties of the medium.
\\

    \item Clumps mass and SFR correlate linearly.
    We found remarkably small scatter across the simulations after averaging over time. As all the samples (of runs without outflows) share the same slope and offset, we can infer that a universal scaling of the SF process at these scales is largely independent of the medium's large-scale properties. This slope is steeper than that inferred semi-analytically for Hi-GAL clumps, but in line with direct observational measures of clump' SFR \citep[e.g.][]{Garcia-Burillo2012, Liu2016}. 
\\

    \item Protostellar outflows increase local fragmentation.
    On average, clumps forming in the \outflowsim tend to be less massive and larger, likely due to enhanced local turbulent support, while the large-radius tail of the $R_\mathrm{sph}$ distribution shows a sharper cut-off. 
\\

    \item Protostellar outflows quench the SFR, but do not affect the $\simuSFR-M$  relation.
    Although decreasing the global $\simuSFR$ by a factor of 4 and acting similarly on single clumps, protostellar outflows do not affect the slope of the mass-SFR relation. Studying the effect on $\eff$, we also found that the agreement with the expected $\eff$ from Hi-GAL clumps decreases, possibly suggesting an overrepresentation of the phenomenon, probably due to either excessive mass ejection or the resolution limitations of the present simulations. 
\\
    
    \item  With a simple scaling argument we recovered the correlation between mass and $\simuSFR$.    Considering three different accretion regimes and assuming a steady state between clump accretion and the $\simuSFR$, a nearly-linear $M-\simuSFR$ relationship arises naturally -- specifically $M^{5/6}$ and $M^{7/9}$ in the turbulent modes, and exactly linear in the gravity-driven case. For $M\approx100\,\MSun$ clumps, these reproduce the simulated $\simuSFR$ within a factor of $\sim 2-3$. 
\\
    \item The clump mass--SFR relation hints at a gravity-regulated mode of SF.
    Although the fitted slope of the $\eff(\tau_\star)\sim M^\eta$ relation cannot definitively distinguish between the supersonic accretion and gravity-dominated regimes, the linear scaling of SFR with mass found in the simulated clump population favours the latter interpretation (Eq.~\ref{eq: accretion models with coeff}). This is further supported by the shallower slope found when a lower $n_\mathrm{thr}$ is adopted (Appendix~\ref{app: HOP thresholds}), possibly indicating a transition to a turbulence-regulated regime.
\\
    \item We found possible hints of a transition between the two regimes of accretion onto the clumps.
    Lowering the \texttt{HOP} $n_\mathrm{thr}$ to $\approx100\,\ndens$ (Appendix~\ref{app: HOP thresholds}), we observe that the slope in the $M-\simuSFR$ tends to flatten, as expected from turbulence-driven accretion modes. 
\end{itemize}
These results emphasise the need for a multi-scale approach in order to disentangle the problem of SF. In future work, zoom-in simulations of selected star-forming regions will provide further insights into reaching dynamical ranges below those achieved here (Abboudeh, in prep.), resolving the smaller-scale dynamics that set the stage for the formation of single stars while preserving the link to the large-scale environment.

\begin{acknowledgements}
We thank the anonymous referee for their comments and constructive feedback that greatly improved the manuscript.
This research has received funding from the European Research Council synergy grant ECOGAL (Grant: 855130).
\end{acknowledgements}

\bibliographystyle{aa} 
\bibliography{aa61450-26}

@ARTICLE{Teyssier2002,
       author = {{Teyssier}, R.},
        title = "{Cosmological hydrodynamics with adaptive mesh refinement. A new high resolution code called RAMSES}",
      journal = {\aap},
         year = 2002,
        month = apr,
       volume = {385},
        pages = {337-364},
          doi = {10.1051/0004-6361:20011817},
archivePrefix = {arXiv},
       eprint = {astro-ph/0111367},
 primaryClass = {astro-ph},
       adsurl = {https://ui.adsabs.harvard.edu/abs/2002A&A...385..337T}
}

@ARTICLE{Barnes2017,
       author = {{Barnes}, A.~T. and {Longmore}, S.~N. and {Battersby}, C. and {Bally}, J. and {Kruijssen}, J.~M.~D. and {Henshaw}, J.~D. and {Walker}, D.~L.},
        title = "{Star formation rates and efficiencies in the Galactic Centre}",
      journal = {\mnras},
         year = 2017,
        month = aug,
       volume = {469},
       number = {2},
        pages = {2263-2285},
          doi = {10.1093/mnras/stx941},
archivePrefix = {arXiv},
       eprint = {1704.03572},
 primaryClass = {astro-ph.GA},
       adsurl = {https://ui.adsabs.harvard.edu/abs/2017MNRAS.469.2263B}
}

@ARTICLE{Zinnecker2007,
       author = {{Zinnecker}, Hans and {Yorke}, Harold W.},
        title = "{Toward Understanding Massive Star Formation}",
      journal = {\araa},
         year = 2007,
        month = sep,
       volume = {45},
       number = {1},
        pages = {481-563},
          doi = {10.1146/annurev.astro.44.051905.092549},
archivePrefix = {arXiv},
       eprint = {0707.1279},
 primaryClass = {astro-ph},
       adsurl = {https://ui.adsabs.harvard.edu/abs/2007ARA&A..45..481Z}
}

@ARTICLE{Lebreuilly2025,
       author = {{Lebreuilly}, Ugo and {Traficante}, Alessio and {Nucara}, Alice and {Tung}, Ngo-Duy and {Hennebelle}, Patrick and {Molinari}, Sergio and {Klessen}, Ralf S. and {Testi}, Leonardo and {Pelkonen}, Veli-Matti and {Benedettini}, Milena and {Coletta}, Alessandro and {Elia}, Davide and {Mininni}, Chiara and {Pezzuto}, Stefania and {Soler}, Juan D. and {Suin}, Paolo and {Toci}, Claudia},
        title = "{The Rosetta Stone Project: I. A suite of radiative magnetohydrodynamics simulations of high-mass star-forming clumps}",
      journal = {\aap},
         year = 2025,
        month = sep,
       volume = {701},
          eid = {A217},
        pages = {A217},
          doi = {10.1051/0004-6361/202554774},
archivePrefix = {arXiv},
       eprint = {2507.08436},
 primaryClass = {astro-ph.SR},
       adsurl = {https://ui.adsabs.harvard.edu/abs/2025A&A...701A.217L}
}

@ARTICLE{Molinari2010,
       author = {{Molinari}, S. and {Swinyard}, B. and {Bally}, J. and {Barlow}, M. and {Bernard}, J.-P. and {Martin}, P. and {Moore}, T. and {Noriega-Crespo}, A. and {Plume}, R. and {Testi}, L. and {Zavagno}, A. and {Abergel}, A. and {Ali}, B. and {Andr{\'e}}, P. and {Baluteau}, J.-P. and {Benedettini}, M. and {Bern{\'e}}, O. and {Billot}, N.~P. and {Blommaert}, J. and {Bontemps}, S. and {Boulanger}, F. and {Brand}, J. and {Brunt}, C. and {Burton}, M. and {Campeggio}, L. and {Carey}, S. and {Caselli}, P. and {Cesaroni}, R. and {Cernicharo}, J. and {Chakrabarti}, S. and {Chrysostomou}, A. and {Codella}, C. and {Cohen}, M. and {Compiegne}, M. and {Davis}, C.~J. and {de Bernardis}, P. and {de Gasperis}, G. and {Di Francesco}, J. and {di Giorgio}, A.~M. and {Elia}, D. and {Faustini}, F. and {Fischera}, J.~F. and {Fukui}, Y. and {Fuller}, G.~A. and {Ganga}, K. and {Garcia-Lario}, P. and {Giard}, M. and {Giardino}, G. and {Glenn}, J. and {Goldsmith}, P. and {Griffin}, M. and {Hoare}, M. and {Huang}, M. and {Jiang}, B. and {Joblin}, C. and {Joncas}, G. and {Juvela}, M. and {Kirk}, J. and {Lagache}, G. and {Li}, J.~Z. and {Lim}, T.~L. and {Lord}, S.~D. and {Lucas}, P.~W. and {Maiolo}, B. and {Marengo}, M. and {Marshall}, D. and {Masi}, S. and {Massi}, F. and {Matsuura}, M. and {Meny}, C. and {Minier}, V. and {Miville-Desch{\^e}nes}, M.-A. and {Montier}, L. and {Motte}, F. and {M{\"u}ller}, T.~G. and {Natoli}, P. and {Neves}, J. and {Olmi}, L. and {Paladini}, R. and {Paradis}, D. and {Pestalozzi}, M. and {Pezzuto}, S. and {Piacentini}, F. and {Pomar{\`e}s}, M. and {Popescu}, C.~C. and {Reach}, W.~T. and {Richer}, J. and {Ristorcelli}, I. and {Roy}, A. and {Royer}, P. and {Russeil}, D. and {Saraceno}, P. and {Sauvage}, M. and {Schilke}, P. and {Schneider-Bontemps}, N. and {Schuller}, F. and {Schultz}, B. and {Shepherd}, D.~S. and {Sibthorpe}, B. and {Smith}, H.~A. and {Smith}, M.~D. and {Spinoglio}, L. and {Stamatellos}, D. and {Strafella}, F. and {Stringfellow}, G. and {Sturm}, E. and {Taylor}, R. and {Thompson}, M.~A. and {Tuffs}, R.~J. and {Umana}, G. and {Valenziano}, L. and {Vavrek}, R. and {Viti}, S. and {Waelkens}, C. and {Ward-Thompson}, D. and {White}, G. and {Wyrowski}, F. and {Yorke}, H.~W. and {Zhang}, Q.},
        title = "{Hi-GAL: The Herschel Infrared Galactic Plane Survey}",
      journal = {\pasp},
         year = 2010,
        month = mar,
       volume = {122},
       number = {889},
        pages = {314},
          doi = {10.1086/651314},
archivePrefix = {arXiv},
       eprint = {1001.2106},
 primaryClass = {astro-ph.GA},
       adsurl = {https://ui.adsabs.harvard.edu/abs/2010PASP..122..314M}
}

@ARTICLE{Traficante2015,
       author = {{Traficante}, A. and {Fuller}, G.~A. and {Pineda}, J.~E. and {Pezzuto}, S.},
        title = "{Hyper: Hybrid photometry and extraction routine}",
      journal = {\aap},
         year = 2015,
        month = feb,
       volume = {574},
          eid = {A119},
        pages = {A119},
          doi = {10.1051/0004-6361/201323254},
archivePrefix = {arXiv},
       eprint = {1410.7293},
 primaryClass = {astro-ph.IM},
       adsurl = {https://ui.adsabs.harvard.edu/abs/2015A&A...574A.119T}
}

@ARTICLE{Brucy2023,
       author = {{Brucy}, No{\'e} and {Hennebelle}, Patrick and {Colman}, Tine and {Iteanu}, Simon},
        title = "{Large-scale turbulent driving regulates star formation in high-redshift gas-rich galaxies. II. Influence of the magnetic field and the turbulent compressive fraction}",
      journal = {\aap},
         year = 2023,
        month = jul,
       volume = {675},
          eid = {A144},
        pages = {A144},
          doi = {10.1051/0004-6361/202244915},
archivePrefix = {arXiv},
       eprint = {2305.18012},
 primaryClass = {astro-ph.GA},
       adsurl = {https://ui.adsabs.harvard.edu/abs/2023A&A...675A.144B}
}

@ARTICLE{Bleuler2014,
       author = {{Bleuler}, Andreas and {Teyssier}, Romain},
        title = "{Towards a more realistic sink particle algorithm for the RAMSES CODE}",
      journal = {\mnras},
         year = 2014,
        month = dec,
       volume = {445},
       number = {4},
        pages = {4015-4036},
          doi = {10.1093/mnras/stu2005},
archivePrefix = {arXiv},
       eprint = {1409.6528},
 primaryClass = {astro-ph.SR},
       adsurl = {https://ui.adsabs.harvard.edu/abs/2014MNRAS.445.4015B}
}

@unpublished{Hennebelle2026,
  author = {Hennebelle, Patrick and Suin, Paolo and Abboudeh, Georges and Schisano, Eugenio and Elia, Davide and Molinari, Sergio and Brucy, No{\'e} and Colman, Tine and Lebreuilly, Ugo and Soler, Juan D. and Nucara, Alice and Pelkonen, Veli-Matti and Traficante, Alessio and Klessen, Ralf and Testi, Leonardo},
  title  = {Title of the paper},
  note   = {submitted to Astronomy \& Astrophysics},
  year   = {2026}
}

@ARTICLE{Urquhart2022,
       author = {{Urquhart}, J.~S. and {Wells}, M.~R.~A. and {Pillai}, T. and {Leurini}, S. and {Giannetti}, A. and {Moore}, T.~J.~T. and {Thompson}, M.~A. and {Figura}, C. and {Colombo}, D. and {Yang}, A.~Y. and {K{\"o}nig}, C. and {Wyrowski}, F. and {Menten}, K.~M. and {Rigby}, A.~J. and {Eden}, D.~J. and {Ragan}, S.~E.},
        title = "{ATLASGAL - evolutionary trends in high-mass star formation}",
      journal = {\mnras},
         year = 2022,
        month = mar,
       volume = {510},
       number = {3},
        pages = {3389-3407},
          doi = {10.1093/mnras/stab3511},
archivePrefix = {arXiv},
       eprint = {2111.12816},
 primaryClass = {astro-ph.GA},
       adsurl = {https://ui.adsabs.harvard.edu/abs/2022MNRAS.510.3389U}
}

@ARTICLE{Priestley2026,
       author = {{Priestley}, F.~D. and {Clark}, P.~C. and {Glover}, S.~C.~O. and {Ragan}, S.~E. and {Stuber}, S.~K. and {Cusack}, M.~T. and {Feh{\'e}r}, O. and {Klessen}, R.~S.},
        title = "{NEATH V: the relationship between line emission from dense gas tracers and the star formation rate}",
      journal = {\mnras},
         year = 2026,
        month = mar,
          doi = {10.1093/mnras/stag527},
       adsurl = {https://ui.adsabs.harvard.edu/abs/2026MNRAS.tmp..488P}
}

@ARTICLE{Nisini2018,
       author = {{Nisini}, B. and {Antoniucci}, S. and {Alcal{\'a}}, J.~M. and {Giannini}, T. and {Manara}, C.~F. and {Natta}, A. and {Fedele}, D. and {Biazzo}, K.},
        title = "{Connection between jets, winds and accretion in T Tauri stars. The X-shooter view}",
      journal = {\aap},
         year = 2018,
        month = jan,
       volume = {609},
          eid = {A87},
        pages = {A87},
          doi = {10.1051/0004-6361/201730834},
archivePrefix = {arXiv},
       eprint = {1710.05587},
 primaryClass = {astro-ph.SR},
       adsurl = {https://ui.adsabs.harvard.edu/abs/2018A&A...609A..87N}
}

@ARTICLE{Guszejnov2023,
       author = {{Guszejnov}, D{\'a}vid and {Raju}, Aman N. and {Offner}, Stella S.~R. and {Grudi{\'c}}, Michael Y. and {Faucher-Gigu{\`e}re}, Claude-Andr{\'e} and {Hopkins}, Philip F. and {Rosen}, Anna L.},
        title = "{Effects of the environment on the multiplicity properties of stars in the STARFORGE simulations}",
      journal = {\mnras},
         year = 2023,
        month = jan,
       volume = {518},
       number = {3},
        pages = {4693-4712},
          doi = {10.1093/mnras/stac3268},
archivePrefix = {arXiv},
       eprint = {2208.02844},
 primaryClass = {astro-ph.GA},
       adsurl = {https://ui.adsabs.harvard.edu/abs/2023MNRAS.518.4693G}
}

@ARTICLE{Federrath2014, 
       author = {{Federrath}, Christoph and {Schr{\"o}n}, Martin and {Banerjee}, Robi and {Klessen}, Ralf S.},
        title = "{Modeling Jet and Outflow Feedback during Star Cluster Formation}",
      journal = {\apj},
         year = 2014,
        month = aug,
       volume = {790},
       number = {2},
          eid = {128},
        pages = {128},
          doi = {10.1088/0004-637X/790/2/128},
archivePrefix = {arXiv},
       eprint = {1406.3625},
 primaryClass = {astro-ph.SR},
       adsurl = {https://ui.adsabs.harvard.edu/abs/2014ApJ...790..128F}
}

@ARTICLE{Federrath2013,
       author = {{Federrath}, Christoph},
        title = "{The origin of physical variations in the star formation law}",
      journal = {\mnras},
         year = 2013,
        month = dec,
       volume = {436},
       number = {4},
        pages = {3167-3172},
          doi = {10.1093/mnras/stt1799},
archivePrefix = {arXiv},
       eprint = {1307.1467},
 primaryClass = {astro-ph.GA},
       adsurl = {https://ui.adsabs.harvard.edu/abs/2013MNRAS.436.3167F}
}

@ARTICLE{Federrath2012,
       author = {{Federrath}, Christoph and {Klessen}, Ralf S.},
        title = "{The Star Formation Rate of Turbulent Magnetized Clouds: Comparing Theory, Simulations, and Observations}",
      journal = {\apj},
         year = 2012,
        month = dec,
       volume = {761},
       number = {2},
          eid = {156},
        pages = {156},
          doi = {10.1088/0004-637X/761/2/156},
archivePrefix = {arXiv},
       eprint = {1209.2856},
 primaryClass = {astro-ph.SR},
       adsurl = {https://ui.adsabs.harvard.edu/abs/2012ApJ...761..156F}
}

@ARTICLE{Heiderman2010,
       author = {{Heiderman}, Amanda and {Evans}, II, Neal J. and {Allen}, Lori E. and {Huard}, Tracy and {Heyer}, Mark},
        title = "{The Star Formation Rate and Gas Surface Density Relation in the Milky Way: Implications for Extragalactic Studies}",
      journal = {\apj},
         year = 2010,
        month = nov,
       volume = {723},
       number = {2},
        pages = {1019-1037},
          doi = {10.1088/0004-637X/723/2/1019},
archivePrefix = {arXiv},
       eprint = {1009.1621},
 primaryClass = {astro-ph.GA},
       adsurl = {https://ui.adsabs.harvard.edu/abs/2010ApJ...723.1019H}
}

@ARTICLE{Hennebelle2008,
       author = {{Hennebelle}, P. and {Fromang}, S.},
        title = "{Magnetic processes in a collapsing dense core. I. Accretion and ejection}",
      journal = {\aap},
         year = 2008,
        month = jan,
       volume = {477},
       number = {1},
        pages = {9-24},
          doi = {10.1051/0004-6361:20078309},
archivePrefix = {arXiv},
       eprint = {0709.2886},
 primaryClass = {astro-ph},
       adsurl = {https://ui.adsabs.harvard.edu/abs/2008A&A...477....9H}
}

@ARTICLE{Matsushita2017,
       author = {{Matsushita}, Yuko and {Machida}, Masahiro N. and {Sakurai}, Yuya and {Hosokawa}, Takashi},
        title = "{Massive outflows driven by magnetic effects in star-forming clouds with high mass accretion rates}",
      journal = {\mnras},
         year = 2017,
        month = sep,
       volume = {470},
       number = {1},
        pages = {1026-1049},
          doi = {10.1093/mnras/stx893},
archivePrefix = {arXiv},
       eprint = {1704.03185},
 primaryClass = {astro-ph.SR},
       adsurl = {https://ui.adsabs.harvard.edu/abs/2017MNRAS.470.1026M}
}

@ARTICLE{Bacciotti2011,
       author = {{Bacciotti}, F. and {Whelan}, E.~T. and {Alcal{\'a}}, J.~M. and {Nisini}, B. and {Podio}, L. and {Randich}, S. and {Stelzer}, B. and {Cupani}, G.},
        title = "{The First X-shooter Observations of Jets from Young Stars}",
      journal = {\apjl},
         year = 2011,
        month = aug,
       volume = {737},
       number = {2},
          eid = {L26},
        pages = {L26},
          doi = {10.1088/2041-8205/737/2/L26},
archivePrefix = {arXiv},
       eprint = {1106.3860},
 primaryClass = {astro-ph.SR},
       adsurl = {https://ui.adsabs.harvard.edu/abs/2011ApJ...737L..26B}
}

@ARTICLE{Khullar2024,
       author = {{Khullar}, Shivan and {Matzner}, Christopher D. and {Murray}, Norman and {Grudi{\'c}}, Michael Y. and {Guszejnov}, D{\'a}vid and {Wetzel}, Andrew and {Hopkins}, Philip F.},
        title = "{Playing with FIRE: A Galactic Feedback-halting Experiment Challenges Star Formation Rate Theories}",
      journal = {\apj},
         year = 2024,
        month = sep,
       volume = {973},
       number = {1},
          eid = {40},
        pages = {40},
          doi = {10.3847/1538-4357/ad6311},
archivePrefix = {arXiv},
       eprint = {2406.18526},
 primaryClass = {astro-ph.GA},
       adsurl = {https://ui.adsabs.harvard.edu/abs/2024ApJ...973...40K}
}

@ARTICLE{Tress2020,
       author = {{Tress}, Robin G. and {Sormani}, Mattia C. and {Glover}, Simon C.~O. and {Klessen}, Ralf S. and {Battersby}, Cara D. and {Clark}, Paul C. and {Hatchfield}, H. Perry and {Smith}, Rowan J.},
        title = "{Simulations of the Milky Way's central molecular zone - I. Gas dynamics}",
      journal = {\mnras},
         year = 2020,
        month = dec,
       volume = {499},
       number = {3},
        pages = {4455-4478},
          doi = {10.1093/mnras/staa3120},
archivePrefix = {arXiv},
       eprint = {2004.06724},
 primaryClass = {astro-ph.GA},
       adsurl = {https://ui.adsabs.harvard.edu/abs/2020MNRAS.499.4455T}
}

@ARTICLE{Tress2021,
       author = {{Tre{\ss}}, Robin G. and {Sormani}, Mattia C. and {Smith}, Rowan J. and {Glover}, Simon C.~O. and {Klessen}, Ralf S. and {Mac Low}, Mordecai-Mark and {Clark}, Paul and {Duarte-Cabral}, Ana},
        title = "{Simulations of the star-forming molecular gas in an interacting M51-like galaxy: cloud population statistics}",
      journal = {\mnras},
         year = 2021,
        month = aug,
       volume = {505},
       number = {4},
        pages = {5438-5459},
          doi = {10.1093/mnras/stab1683},
archivePrefix = {arXiv},
       eprint = {2012.05919},
 primaryClass = {astro-ph.GA},
       adsurl = {https://ui.adsabs.harvard.edu/abs/2021MNRAS.505.5438T}
}

@ARTICLE{Goller2025,
       author = {{G{\"o}ller}, Junia and {Girichidis}, Philipp and {Brucy}, No{\'e} and {Hunter}, Glen and {Kjellgren}, Karin and {Tress}, Robin and {Klessen}, Ralf S. and {Glover}, Simon C.~O. and {Hennebelle}, Patrick and {Molinari}, Sergio and {Smith}, Rowan and {Soler}, Juan D. and {Sormani}, Mattia C. and {Testi}, Leonardo},
        title = "{Introducing the Rhea simulations of Milky Way-like galaxies: I. Effect of gravitational potential on morphology and star formation}",
      journal = {\aap},
         year = 2025,
        month = dec,
       volume = {704},
          eid = {A331},
        pages = {A331},
          doi = {10.1051/0004-6361/202452223},
archivePrefix = {arXiv},
       eprint = {2502.02646},
 primaryClass = {astro-ph.GA},
       adsurl = {https://ui.adsabs.harvard.edu/abs/2025A&A...704A.331G}
}

@ARTICLE{Sormani2020,
       author = {{Sormani}, Mattia C. and {Tress}, Robin G. and {Glover}, Simon C.~O. and {Klessen}, Ralf S. and {Battersby}, Cara D. and {Clark}, Paul C. and {Hatchfield}, H. Perry and {Smith}, Rowan J.},
        title = "{Simulations of the Milky Way's Central Molecular Zone - II. Star formation}",
      journal = {\mnras},
         year = 2020,
        month = oct,
       volume = {497},
       number = {4},
        pages = {5024-5040},
          doi = {10.1093/mnras/staa1999},
archivePrefix = {arXiv},
       eprint = {2004.06731},
 primaryClass = {astro-ph.GA},
       adsurl = {https://ui.adsabs.harvard.edu/abs/2020MNRAS.497.5024S}
}

@ARTICLE{Girichidis2020,
       author = {{Girichidis}, Philipp and {Offner}, Stella S.~R. and {Kritsuk}, Alexei G. and {Klessen}, Ralf S. and {Hennebelle}, Patrick and {Kruijssen}, J.~M. Diederik and {Krause}, Martin G.~H. and {Glover}, Simon C.~O. and {Padovani}, Marco},
        title = "{Physical Processes in Star Formation}",
      journal = {\ssr},
         year = 2020,
        month = jun,
       volume = {216},
       number = {4},
          eid = {68},
        pages = {68},
          doi = {10.1007/s11214-020-00693-8},
archivePrefix = {arXiv},
       eprint = {2005.06472},
 primaryClass = {astro-ph.GA},
       adsurl = {https://ui.adsabs.harvard.edu/abs/2020SSRv..216...68G}
}

@ARTICLE{Klessen2016,
       author = {{Klessen}, Ralf S. and {Glover}, Simon C.~O.},
        title = "{Physical Processes in the Interstellar Medium}",
      journal = {Saas-Fee Advanced Course},
         year = 2016,
        month = jan,
       volume = {43},
        pages = {85},
          doi = {10.1007/978-3-662-47890-5_2},
archivePrefix = {arXiv},
       eprint = {1412.5182},
 primaryClass = {astro-ph.GA},
       adsurl = {https://ui.adsabs.harvard.edu/abs/2016SAAS...43...85K}
}

@ARTICLE{Brucy2025,
       author = {{Brucy}, No{\'e} and {V{\'a}zquez-Semadeni}, Enrique and {Colman}, Tine and {Fensch}, J{\'e}r{\'e}my and {Klessen}, Ralf S.},
        title = "{What is the contribution of gravitational infall on the mass assembly of star-forming clouds? A case study in a numerical simulation of the interstellar medium}",
      journal = {The Open Journal of Astrophysics},
         year = 2025,
        month = dec,
       volume = {8},
        pages = {54637},
          doi = {10.33232/001c.154637},
archivePrefix = {arXiv},
       eprint = {2510.09480},
 primaryClass = {astro-ph.GA},
       adsurl = {https://ui.adsabs.harvard.edu/abs/2025OJAp....854637B}
}

@ARTICLE{Butler2012,
       author = {{Butler}, Michael J. and {Tan}, Jonathan C.},
        title = "{Mid-infrared Extinction Mapping of Infrared Dark Clouds. II. The Structure of Massive Starless Cores and Clumps}",
      journal = {\apj},
         year = 2012,
        month = jul,
       volume = {754},
       number = {1},
          eid = {5},
        pages = {5},
          doi = {10.1088/0004-637X/754/1/5},
archivePrefix = {arXiv},
       eprint = {1205.2391},
 primaryClass = {astro-ph.SR},
       adsurl = {https://ui.adsabs.harvard.edu/abs/2012ApJ...754....5B}
}

@ARTICLE{Evans2014,
       author = {{Evans}, Neal J., II and {Heiderman}, Amanda and {Vutisalchavakul}, Nalin},
        title = "{Star Formation Relations in Nearby Molecular Clouds}",
      journal = {\apj},
         year = 2014,
        month = feb,
       volume = {782},
       number = {2},
          eid = {114},
        pages = {114},
          doi = {10.1088/0004-637X/782/2/114},
archivePrefix = {arXiv},
       eprint = {1401.3287},
 primaryClass = {astro-ph.GA},
       adsurl = {https://ui.adsabs.harvard.edu/abs/2014ApJ...782..114E}
}

@ARTICLE{Krumholz2014,
       author = {{Krumholz}, Mark R.},
        title = "{The big problems in star formation: The star formation rate, stellar clustering, and the initial mass function}",
      journal = {\physrep},
         year = 2014,
        month = jun,
       volume = {539},
        pages = {49-134},
          doi = {10.1016/j.physrep.2014.02.001},
archivePrefix = {arXiv},
       eprint = {1402.0867},
 primaryClass = {astro-ph.GA},
       adsurl = {https://ui.adsabs.harvard.edu/abs/2014PhR...539...49K}
}

@ARTICLE{Kennicutt2012,
       author = {{Kennicutt}, Robert C. and {Evans}, Neal J.},
        title = "{Star Formation in the Milky Way and Nearby Galaxies}",
      journal = {\araa},
         year = 2012,
        month = sep,
       volume = {50},
        pages = {531-608},
          doi = {10.1146/annurev-astro-081811-125610},
archivePrefix = {arXiv},
       eprint = {1204.3552},
 primaryClass = {astro-ph.GA},
       adsurl = {https://ui.adsabs.harvard.edu/abs/2012ARA&A..50..531K}
}

@ARTICLE{Hoang2025,
       author = {{Hoang}, Thanh Dat and {Lee}, Min-Young and {Wyrowski}, Friedrich and {Karska}, Agata and {Navarete}, Felipe and {Menten}, Karl M.},
        title = "{ATLASGAL-selected high-mass clumps in the inner Galaxy: XI. Morphology and kinematics of warm inner envelopes}",
      journal = {\aap},
         year = 2025,
        month = mar,
       volume = {695},
          eid = {A24},
        pages = {A24},
          doi = {10.1051/0004-6361/202452371},
archivePrefix = {arXiv},
       eprint = {2501.13752},
 primaryClass = {astro-ph.GA},
       adsurl = {https://ui.adsabs.harvard.edu/abs/2025A&A...695A..24H}
}

@ARTICLE{Mueller2002,
       author = {{Mueller}, Kaisa E. and {Shirley}, Yancy L. and {Evans}, II, Neal J. and {Jacobson}, Heather R.},
        title = "{The Physical Conditions for Massive Star Formation: Dust Continuum Maps and Modeling}",
      journal = {\apjs},
         year = 2002,
        month = dec,
       volume = {143},
       number = {2},
        pages = {469-497},
          doi = {10.1086/342881},
archivePrefix = {arXiv},
       eprint = {astro-ph/0207322},
 primaryClass = {astro-ph},
       adsurl = {https://ui.adsabs.harvard.edu/abs/2002ApJS..143..469M}
}

@INPROCEEDINGS{Frank2014,
       author = {{Frank}, A. and {Ray}, T.~P. and {Cabrit}, S. and {Hartigan}, P. and {Arce}, H.~G. and {Bacciotti}, F. and {Bally}, J. and {Benisty}, M. and {Eisl{\"o}ffel}, J. and {G{\"u}del}, M. and {Lebedev}, S. and {Nisini}, B. and {Raga}, A.},
        title = "{Jets and Outflows from Star to Cloud: Observations Confront Theory}",
    booktitle = {Protostars and Planets VI},
         year = 2014,
       editor = {{Beuther}, Henrik and {Klessen}, Ralf S. and {Dullemond}, Cornelis P. and {Henning}, Thomas},
        month = jan,
        pages = {451-474},
          doi = {10.2458/azu_uapress_9780816531240-ch020},
archivePrefix = {arXiv},
       eprint = {1402.3553},
 primaryClass = {astro-ph.SR},
       adsurl = {https://ui.adsabs.harvard.edu/abs/2014prpl.conf..451F}
}

@ARTICLE{Pokhrel2021,
       author = {{Pokhrel}, Riwaj and {Gutermuth}, Robert A. and {Krumholz}, Mark R. and {Federrath}, Christoph and {Heyer}, Mark and {Khullar}, Shivan and {Megeath}, S. Thomas and {Myers}, Philip C. and {Offner}, Stella S.~R. and {Pipher}, Judith L. and {Fischer}, William J. and {Henning}, Thomas and {Hora}, Joseph L.},
        title = "{The Single-cloud Star Formation Relation}",
      journal = {\apjl},
         year = 2021,
        month = may,
       volume = {912},
       number = {1},
          eid = {L19},
        pages = {L19},
          doi = {10.3847/2041-8213/abf564},
archivePrefix = {arXiv},
       eprint = {2104.04551},
 primaryClass = {astro-ph.GA},
       adsurl = {https://ui.adsabs.harvard.edu/abs/2021ApJ...912L..19P}
}

@ARTICLE{Seifried2017,
       author = {{Seifried}, D. and {Walch}, S. and {Girichidis}, P. and {Naab}, T. and {W{\"u}nsch}, R. and {Klessen}, R.~S. and {Glover}, S.~C.~O. and {Peters}, T. and {Clark}, P.},
        title = "{SILCC-Zoom: the dynamic and chemical evolution of molecular clouds}",
      journal = {\mnras},
         year = 2017,
        month = dec,
       volume = {472},
       number = {4},
        pages = {4797-4818},
          doi = {10.1093/mnras/stx2343},
archivePrefix = {arXiv},
       eprint = {1704.06487},
 primaryClass = {astro-ph.GA},
       adsurl = {https://ui.adsabs.harvard.edu/abs/2017MNRAS.472.4797S}
}

@ARTICLE{Smith2008,
       author = {{Smith}, Rowan J. and {Clark}, Paul C. and {Bonnell}, Ian A.},
        title = "{The structure of molecular clouds and the universality of the clump mass function}",
      journal = {\mnras},
         year = 2008,
        month = dec,
       volume = {391},
       number = {3},
        pages = {1091-1099},
          doi = {10.1111/j.1365-2966.2008.13970.x},
archivePrefix = {arXiv},
       eprint = {0809.2702},
 primaryClass = {astro-ph},
       adsurl = {https://ui.adsabs.harvard.edu/abs/2008MNRAS.391.1091S}
}

@ARTICLE{Gomez2014,
       author = {{G{\'o}mez}, Gilberto C. and {V{\'a}zquez-Semadeni}, Enrique},
        title = "{Filaments in Simulations of Molecular Cloud Formation}",
      journal = {\apj},
         year = 2014,
        month = aug,
       volume = {791},
       number = {2},
          eid = {124},
        pages = {124},
          doi = {10.1088/0004-637X/791/2/124},
archivePrefix = {arXiv},
       eprint = {1308.6298},
 primaryClass = {astro-ph.GA},
       adsurl = {https://ui.adsabs.harvard.edu/abs/2014ApJ...791..124G}
}

@ARTICLE{Murray2018,
       author = {{Murray}, Daniel and {Goyal}, Shivam and {Chang}, Philip},
        title = "{The effects of protostellar jet feedback on turbulent collapse}",
      journal = {\mnras},
         year = 2018,
        month = mar,
       volume = {475},
       number = {1},
        pages = {1023-1035},
          doi = {10.1093/mnras/stx3153},
archivePrefix = {arXiv},
       eprint = {1710.09415},
 primaryClass = {astro-ph.GA},
       adsurl = {https://ui.adsabs.harvard.edu/abs/2018MNRAS.475.1023M}
}

@ARTICLE{Appel2023,
       author = {{Appel}, Sabrina M. and {Burkhart}, Blakesley and {Semenov}, Vadim A. and {Federrath}, Christoph and {Rosen}, Anna L. and {Tan}, Jonathan C.},
        title = "{What Sets the Star Formation Rate of Molecular Clouds? The Density Distribution as a Fingerprint of Compression and Expansion Rates}",
      journal = {\apj},
         year = 2023,
        month = sep,
       volume = {954},
       number = {1},
          eid = {93},
        pages = {93},
          doi = {10.3847/1538-4357/ace897},
archivePrefix = {arXiv},
       eprint = {2301.07723},
 primaryClass = {astro-ph.GA},
       adsurl = {https://ui.adsabs.harvard.edu/abs/2023ApJ...954...93A}
}

@ARTICLE{Federrath2015,
       author = {{Federrath}, Christoph},
        title = "{Inefficient star formation through turbulence, magnetic fields and feedback}",
      journal = {\mnras},
         year = 2015,
        month = jul,
       volume = {450},
       number = {4},
        pages = {4035-4042},
          doi = {10.1093/mnras/stv941},
archivePrefix = {arXiv},
       eprint = {1504.03690},
 primaryClass = {astro-ph.SR},
       adsurl = {https://ui.adsabs.harvard.edu/abs/2015MNRAS.450.4035F}
}

@ARTICLE{Suin2024,
       author = {{Suin}, P. and {Zavagno}, A. and {Colman}, T. and {Hennebelle}, P. and {Verliat}, A. and {Russeil}, D.},
        title = "{Stellar feedback in the star formation-gas density relation: Comparison between simulations and observations}",
      journal = {\aap},
         year = 2024,
        month = feb,
       volume = {682},
          eid = {A76},
        pages = {A76},
          doi = {10.1051/0004-6361/202347527},
archivePrefix = {arXiv},
       eprint = {2311.18522},
 primaryClass = {astro-ph.GA},
       adsurl = {https://ui.adsabs.harvard.edu/abs/2024A&A...682A..76S}
}

@ARTICLE{Keller2022,
       author = {{Keller}, Benjamin W. and {Kruijssen}, J.~M. Diederik and {Chevance}, M{\'e}lanie},
        title = "{Empirically motivated early feedback: momentum input by stellar feedback in galaxy simulations inferred through observations}",
      journal = {\mnras},
         year = 2022,
        month = aug,
       volume = {514},
       number = {4},
        pages = {5355-5374},
          doi = {10.1093/mnras/stac1607},
archivePrefix = {arXiv},
       eprint = {2206.06391},
 primaryClass = {astro-ph.GA},
       adsurl = {https://ui.adsabs.harvard.edu/abs/2022MNRAS.514.5355K}
}

@ARTICLE{Smith2020,
       author = {{Smith}, Rowan J. and {Tre{\ss}}, Robin G. and {Sormani}, Mattia C. and {Glover}, Simon C.~O. and {Klessen}, Ralf S. and {Clark}, Paul C. and {Izquierdo}, Andr{\'e}s F. and {Duarte-Cabral}, Ana and {Zucker}, Catherine},
        title = "{The Cloud Factory I: Generating resolved filamentary molecular clouds from galactic-scale forces}",
      journal = {\mnras},
         year = 2020,
        month = feb,
       volume = {492},
       number = {2},
        pages = {1594-1613},
          doi = {10.1093/mnras/stz3328},
archivePrefix = {arXiv},
       eprint = {1911.05753},
 primaryClass = {astro-ph.GA},
       adsurl = {https://ui.adsabs.harvard.edu/abs/2020MNRAS.492.1594S}
}

@ARTICLE{Feng2024,
       author = {{Feng}, Jiancheng and {Smith}, Rowan J. and {Hacar}, Alvaro and {Clark}, Susan E. and {Seifried}, Daniel},
        title = "{On the evolution of the observed mass-to-length relationship for star-forming filaments}",
      journal = {\mnras},
         year = 2024,
        month = mar,
       volume = {528},
       number = {4},
        pages = {6370-6387},
          doi = {10.1093/mnras/stae407},
archivePrefix = {arXiv},
       eprint = {2402.05186},
 primaryClass = {astro-ph.GA},
       adsurl = {https://ui.adsabs.harvard.edu/abs/2024MNRAS.528.6370F}
}

@ARTICLE{Grudic2021,
       author = {{Grudi{\'c}}, Michael Y. and {Guszejnov}, D{\'a}vid and {Hopkins}, Philip F. and {Offner}, Stella S.~R. and {Faucher-Gigu{\`e}re}, Claude-Andr{\'e}},
        title = "{STARFORGE: Towards a comprehensive numerical model of star cluster formation and feedback}",
      journal = {\mnras},
         year = 2021,
        month = sep,
       volume = {506},
       number = {2},
        pages = {2199-2231},
          doi = {10.1093/mnras/stab1347},
archivePrefix = {arXiv},
       eprint = {2010.11254},
 primaryClass = {astro-ph.IM},
       adsurl = {https://ui.adsabs.harvard.edu/abs/2021MNRAS.506.2199G}
}

@ARTICLE{Renaud2024,
       author = {{Renaud}, Florent and {Agertz}, Oscar and {Romeo}, Alessandro B.},
        title = "{From giant clumps to clouds. IV. Extreme star-forming clumps on top of universal cloud scaling relations in gas-rich galaxies}",
      journal = {\aap},
         year = 2024,
        month = jul,
       volume = {687},
          eid = {A91},
        pages = {A91},
          doi = {10.1051/0004-6361/202449721},
archivePrefix = {arXiv},
       eprint = {2402.15557},
 primaryClass = {astro-ph.GA},
       adsurl = {https://ui.adsabs.harvard.edu/abs/2024A&A...687A..91R}
}

@ARTICLE{Colman2025,
       author = {{Colman}, Tine and {Hennebelle}, Patrick and {Brucy}, Noe and {Dumond}, Pierre and {Girichidis}, Philipp and {Soler}, Juan and {Glover}, Simon and {Klessen}, Ralf and {Miville-Deschenes}, Marc-Antoine and {Traficante}, Alessio and {Molinari}, Sergio and {Smith}, Rowan and {Testi}, Leonardo},
        title = "{The role of turbulence in setting the phase of the ISM and implications for the star formation rate}",
      journal = {The Open Journal of Astrophysics},
         year = 2025,
        month = nov,
       volume = {8},
        pages = {47517},
          doi = {10.33232/001c.147517},
archivePrefix = {arXiv},
       eprint = {2503.03305},
 primaryClass = {astro-ph.GA},
       adsurl = {https://ui.adsabs.harvard.edu/abs/2025OJAp....847517C}
}

@ARTICLE{Battersby2017,
       author = {{Battersby}, Cara and {Bally}, John and {Svoboda}, Brian},
        title = "{The Lifetimes of Phases in High-mass Star-forming Regions}",
      journal = {\apj},
         year = 2017,
        month = feb,
       volume = {835},
       number = {2},
          eid = {263},
        pages = {263},
          doi = {10.3847/1538-4357/835/2/263},
archivePrefix = {arXiv},
       eprint = {1702.02199},
 primaryClass = {astro-ph.GA},
       adsurl = {https://ui.adsabs.harvard.edu/abs/2017ApJ...835..263B}
}

@ARTICLE{Urquhart2018,
       author = {{Urquhart}, J.~S. and {K{\"o}nig}, C. and {Giannetti}, A. and {Leurini}, S. and {Moore}, T.~J.~T. and {Eden}, D.~J. and {Pillai}, T. and {Thompson}, M.~A. and {Braiding}, C. and {Burton}, M.~G. and {Csengeri}, T. and {Dempsey}, J.~T. and {Figura}, C. and {Froebrich}, D. and {Menten}, K.~M. and {Schuller}, F. and {Smith}, M.~D. and {Wyrowski}, F.},
        title = "{ATLASGAL - properties of a complete sample of Galactic clumps}",
      journal = {\mnras},
         year = 2018,
        month = jan,
       volume = {473},
       number = {1},
        pages = {1059-1102},
          doi = {10.1093/mnras/stx2258},
archivePrefix = {arXiv},
       eprint = {1709.00392},
 primaryClass = {astro-ph.GA},
       adsurl = {https://ui.adsabs.harvard.edu/abs/2018MNRAS.473.1059U}
}

@ARTICLE{Baldeschi2017,
       author = {{Baldeschi}, Adriano and {Elia}, D. and {Molinari}, S. and {Pezzuto}, S. and {Schisano}, E. and {Gatti}, M. and {Serra}, A. and {Merello}, M. and {Benedettini}, M. and {Di Giorgio}, A.~M. and {Liu}, J.~S.},
        title = "{Distance biases in the estimation of the physical properties of Hi-GAL compact sources - I. Clump properties and the identification of high-mass star-forming candidates}",
      journal = {\mnras},
         year = 2017,
        month = apr,
       volume = {466},
       number = {3},
        pages = {3682-3705},
          doi = {10.1093/mnras/stw3353},
archivePrefix = {arXiv},
       eprint = {1701.08035},
 primaryClass = {astro-ph.GA},
       adsurl = {https://ui.adsabs.harvard.edu/abs/2017MNRAS.466.3682B}
}

@ARTICLE{Garcia-Burillo2012,
       author = {{Garc{\'\i}a-Burillo}, S. and {Usero}, A. and {Alonso-Herrero}, A. and {Graci{\'a}-Carpio}, J. and {Pereira-Santaella}, M. and {Colina}, L. and {Planesas}, P. and {Arribas}, S.},
        title = "{Star-formation laws in luminous infrared galaxies. New observational constraints on models}",
      journal = {\aap},
         year = 2012,
        month = mar,
       volume = {539},
          eid = {A8},
        pages = {A8},
          doi = {10.1051/0004-6361/201117838},
archivePrefix = {arXiv},
       eprint = {1111.6773},
 primaryClass = {astro-ph.CO},
       adsurl = {https://ui.adsabs.harvard.edu/abs/2012A&A...539A...8G}
}

@ARTICLE{Gallagher2018,
       author = {{Gallagher}, Molly J. and {Leroy}, Adam K. and {Bigiel}, Frank and {Cormier}, Diane and {Jim{\'e}nez-Donaire}, Mar{\'\i}a J. and {Ostriker}, Eve and {Usero}, Antonio and {Bolatto}, Alberto D. and {Garc{\'\i}a-Burillo}, Santiago and {Hughes}, Annie and {Kepley}, Amanda A. and {Krumholz}, Mark and {Meidt}, Sharon E. and {Meier}, David S. and {Murphy}, Eric J. and {Pety}, J{\'e}r{\^o}me and {Rosolowsky}, Erik and {Schinnerer}, Eva and {Schruba}, Andreas and {Walter}, Fabian},
        title = "{Dense Gas, Dynamical Equilibrium Pressure, and Star Formation in Nearby Star-forming Galaxies}",
      journal = {\apj},
         year = 2018,
        month = may,
       volume = {858},
       number = {2},
          eid = {90},
        pages = {90},
          doi = {10.3847/1538-4357/aabad8},
archivePrefix = {arXiv},
       eprint = {1803.10785},
 primaryClass = {astro-ph.GA},
       adsurl = {https://ui.adsabs.harvard.edu/abs/2018ApJ...858...90G}
}

@ARTICLE{Liu2016,
       author = {{Liu}, Tie and {Kim}, Kee-Tae and {Yoo}, Hyunju and {Liu}, Sheng-yuan and {Tatematsu}, Ken'ichi and {Qin}, Sheng-Li and {Zhang}, Qizhou and {Wu}, Yuefang and {Wang}, Ke and {Goldsmith}, Paul F. and et al.},
        title = "{Star Formation Laws in Both Galactic Massive Clumps and External Galaxies: Extensive Study with Dust Coninuum, HCN (4-3), and CS (7-6)}",
      journal = {\apj},
         year = 2016,
        month = oct,
       volume = {829},
       number = {2},
          eid = {59},
        pages = {59},
          doi = {10.3847/0004-637X/829/2/59},
archivePrefix = {arXiv},
       eprint = {1605.08979},
 primaryClass = {astro-ph.GA},
       adsurl = {https://ui.adsabs.harvard.edu/abs/2016ApJ...829...59L}
}

@ARTICLE{Stephens2016,
       author = {{Stephens}, Ian W. and {Jackson}, James M. and {Whitaker}, J. Scott and {Contreras}, Yanett and {Guzm{\'a}n}, Andr{\'e}s E. and {Sanhueza}, Patricio and {Foster}, Jonathan B. and {Rathborne}, Jill M.},
        title = "{Linking Dense Gas from the Milky Way to External Galaxies}",
      journal = {\apj},
         year = 2016,
        month = jun,
       volume = {824},
       number = {1},
          eid = {29},
        pages = {29},
          doi = {10.3847/0004-637X/824/1/29},
archivePrefix = {arXiv},
       eprint = {1603.09341},
 primaryClass = {astro-ph.GA},
       adsurl = {https://ui.adsabs.harvard.edu/abs/2016ApJ...824...29S}
}

@ARTICLE{Jimenez-Donaire2019,
       author = {{Jim{\'e}nez-Donaire}, Mar{\'\i}a J. and {Bigiel}, F. and {Leroy}, A.~K. and {Usero}, A. and {Cormier}, D. and {Puschnig}, J. and {Gallagher}, M. and {Kepley}, A. and {Bolatto}, A. and {Garc{\'\i}a-Burillo}, S. and {Hughes}, A. and {Kramer}, C. and {Pety}, J. and {Schinnerer}, E. and {Schruba}, A. and {Schuster}, K. and {Walter}, F.},
        title = "{EMPIRE: The IRAM 30 m Dense Gas Survey of Nearby Galaxies}",
      journal = {\apj},
         year = 2019,
        month = aug,
       volume = {880},
       number = {2},
          eid = {127},
        pages = {127},
          doi = {10.3847/1538-4357/ab2b95},
archivePrefix = {arXiv},
       eprint = {1906.08779},
 primaryClass = {astro-ph.GA},
       adsurl = {https://ui.adsabs.harvard.edu/abs/2019ApJ...880..127J}
}

@ARTICLE{Neumann2025,
       author = {{Neumann}, Lukas and {Jim{\'e}nez-Donaire}, Mar{\'\i}a J. and {Leroy}, Adam K. and {Bigiel}, Frank and {Usero}, Antonio and {Sun}, Jiayi and {Schinnerer}, Eva and {Querejeta}, Miguel and {Stuber}, Sophia K. and {Be{\v{s}}li{\'c}}, Ivana and {Barnes}, Ashley and {den Brok}, Jakob and {Cao}, Yixian and {Eibensteiner}, Cosima and {He}, Hao and {Klessen}, Ralf S. and {Liang}, Fu-Heng and {Liu}, Daizhong and {Pan}, Hsi-An and {Williams}, Thomas G.},
        title = "{Dense gas scaling relations at kiloparsec scales across nearby galaxies with the ALMA ALMOND and IRAM 30 m EMPIRE surveys}",
      journal = {\aap},
         year = 2025,
        month = jan,
       volume = {693},
          eid = {L13},
        pages = {L13},
          doi = {10.1051/0004-6361/202453208},
archivePrefix = {arXiv},
       eprint = {2412.10506},
 primaryClass = {astro-ph.GA},
       adsurl = {https://ui.adsabs.harvard.edu/abs/2025A&A...693L..13N}
}

@ARTICLE{Bigiel2008,
       author = {{Bigiel}, F. and {Leroy}, A. and {Walter}, F. and {Brinks}, E. and {de Blok}, W.~J.~G. and {Madore}, B. and {Thornley}, M.~D.},
        title = "{The Star Formation Law in Nearby Galaxies on Sub-Kpc Scales}",
      journal = {\aj},
         year = 2008,
        month = dec,
       volume = {136},
       number = {6},
        pages = {2846-2871},
          doi = {10.1088/0004-6256/136/6/2846},
archivePrefix = {arXiv},
       eprint = {0810.2541},
 primaryClass = {astro-ph},
       adsurl = {https://ui.adsabs.harvard.edu/abs/2008AJ....136.2846B}
}

@ARTICLE{Gao2004,
       author = {{Gao}, Yu and {Solomon}, Philip M.},
        title = "{The Star Formation Rate and Dense Molecular Gas in Galaxies}",
      journal = {\apj},
         year = 2004,
        month = may,
       volume = {606},
       number = {1},
        pages = {271-290},
          doi = {10.1086/382999},
archivePrefix = {arXiv},
       eprint = {astro-ph/0310339},
 primaryClass = {astro-ph},
       adsurl = {https://ui.adsabs.harvard.edu/abs/2004ApJ...606..271G}
}

@ARTICLE{Wu2005,
       author = {{Wu}, Jingwen and {Evans}, II, Neal J. and {Gao}, Yu and {Solomon}, Philip M. and {Shirley}, Yancy L. and {Vanden Bout}, Paul A.},
        title = "{Connecting Dense Gas Tracers of Star Formation in our Galaxy to High-z Star Formation}",
      journal = {\apjl},
         year = 2005,
        month = dec,
       volume = {635},
       number = {2},
        pages = {L173-L176},
          doi = {10.1086/499623},
archivePrefix = {arXiv},
       eprint = {astro-ph/0511424},
 primaryClass = {astro-ph},
       adsurl = {https://ui.adsabs.harvard.edu/abs/2005ApJ...635L.173W}
}

@ARTICLE{Lada2010,
       author = {{Lada}, Charles J. and {Lombardi}, Marco and {Alves}, Jo{\~a}o F.},
        title = "{On the Star Formation Rates in Molecular Clouds}",
      journal = {\apj},
         year = 2010,
        month = nov,
       volume = {724},
       number = {1},
        pages = {687-693},
          doi = {10.1088/0004-637X/724/1/687},
archivePrefix = {arXiv},
       eprint = {1009.2985},
 primaryClass = {astro-ph.GA},
       adsurl = {https://ui.adsabs.harvard.edu/abs/2010ApJ...724..687L}
}

@ARTICLE{Eden2024,
       author = {{Eden}, D.~J. and {Liu}, Tie and {Moore}, T.~J.~T. and {Di Francesco}, J. and {Fuller}, G. and {Kim}, Kee-Tae and {Li}, Di and {Liu}, S.-Y. and {Plume}, R. and {Tatematsu}, Ken'ichi and {Thompson}, M.~A. and {Wu}, Y. and {Bronfman}, L. and {Butner}, H.~M. and {Currie}, M.~J. and {Garay}, G. and {Goldsmith}, P.~F. and {Hirano}, N. and {Johnstone}, D. and {Juvela}, M. and {Lai}, S.-P. and {Lee}, C.~W. and {Mannfors}, E.~E. and {Olguin}, F. and {Pattle}, K. and {Park}, Geumsook and {Polychroni}, D. and {Rawlings}, M. and {Rigby}, A.~J. and {Sanhueza}, P. and {Traficante}, A. and {Urquhart}, J.~S. and {Weferling}, B. and {White}, G.~J. and {Yadav}, R.~K.},
        title = "{A study of Galactic Plane Planck Galactic cold clumps observed by SCOPE and the JCMT Plane Survey}",
      journal = {\mnras},
         year = 2024,
        month = jun,
       volume = {530},
       number = {4},
        pages = {5192-5208},
          doi = {10.1093/mnras/stae1179},
archivePrefix = {arXiv},
       eprint = {2405.00493},
 primaryClass = {astro-ph.GA},
       adsurl = {https://ui.adsabs.harvard.edu/abs/2024MNRAS.530.5192E}
}

@ARTICLE{Eden2017,
       author = {{Eden}, D.~J. and {Moore}, T.~J.~T. and {Plume}, R. and {Urquhart}, J.~S. and {Thompson}, M.~A. and {Parsons}, H. and {Dempsey}, J.~T. and {Rigby}, A.~J. and {Morgan}, L.~K. and {Thomas}, H.~S. and {Berry}, D. and {Buckle}, J. and {Brunt}, C.~M. and {Butner}, H.~M. and {Carretero}, D. and {Chrysostomou}, A. and {Currie}, M.~J. and {deVilliers}, H.~M. and {Fich}, M. and {Gibb}, A.~G. and {Hoare}, M.~G. and {Jenness}, T. and {Manser}, G. and {Mottram}, J.~C. and {Natario}, C. and {Olguin}, F. and {Peretto}, N. and {Pestalozzi}, M. and {Polychroni}, D. and {Redman}, R.~O. and {Salji}, C. and {Summers}, L.~J. and {Tahani}, K. and {Traficante}, A. and {diFrancesco}, J. and {Evans}, A. and {Fuller}, G.~A. and {Johnstone}, D. and {Joncas}, G. and {Longmore}, S.~N. and {Martin}, P.~G. and {Richer}, J.~S. and {Weferling}, B. and {White}, G.~J. and {Zhu}, M.},
        title = "{The JCMT Plane Survey: first complete data release - emission maps and compact source catalogue}",
      journal = {\mnras},
         year = 2017,
        month = aug,
       volume = {469},
       number = {2},
        pages = {2163-2183},
          doi = {10.1093/mnras/stx874},
archivePrefix = {arXiv},
       eprint = {1704.02982},
 primaryClass = {astro-ph.GA},
       adsurl = {https://ui.adsabs.harvard.edu/abs/2017MNRAS.469.2163E}
}

@ARTICLE{Schuller2009,
       author = {{Schuller}, F. and {Menten}, K.~M. and {Contreras}, Y. and {Wyrowski}, F. and {Schilke}, P. and {Bronfman}, L. and {Henning}, T. and {Walmsley}, C.~M. and {Beuther}, H. and {Bontemps}, S. and {Cesaroni}, R. and {Deharveng}, L. and {Garay}, G. and {Herpin}, F. and {Lefloch}, B. and {Linz}, H. and {Mardones}, D. and {Minier}, V. and {Molinari}, S. and {Motte}, F. and {Nyman}, L.-{\r{A}}. and {Reveret}, V. and {Risacher}, C. and {Russeil}, D. and {Schneider}, N. and {Testi}, L. and {Troost}, T. and {Vasyunina}, T. and {Wienen}, M. and {Zavagno}, A. and {Kovacs}, A. and {Kreysa}, E. and {Siringo}, G. and {Wei{\ss}}, A.},
        title = "{ATLASGAL - The APEX telescope large area survey of the galaxy at 870 {\ensuremath{\mu}}m}",
      journal = {\aap},
         year = 2009,
        month = sep,
       volume = {504},
       number = {2},
        pages = {415-427},
          doi = {10.1051/0004-6361/200811568},
archivePrefix = {arXiv},
       eprint = {0903.1369},
 primaryClass = {astro-ph.GA},
       adsurl = {https://ui.adsabs.harvard.edu/abs/2009A&A...504..415S}
}

@ARTICLE{Urquhart2014,
       author = {{Urquhart}, J.~S. and {Moore}, T.~J.~T. and {Csengeri}, T. and {Wyrowski}, F. and {Schuller}, F. and {Hoare}, M.~G. and {Lumsden}, S.~L. and {Mottram}, J.~C. and {Thompson}, M.~A. and {Menten}, K.~M. and {Walmsley}, C.~M. and {Bronfman}, L. and {Pfalzner}, S. and {K{\"o}nig}, C. and {Wienen}, M.},
        title = "{ATLASGAL - towards a complete sample of massive star forming clumps}",
      journal = {\mnras},
         year = 2014,
        month = sep,
       volume = {443},
       number = {2},
        pages = {1555-1586},
          doi = {10.1093/mnras/stu1207},
archivePrefix = {arXiv},
       eprint = {1406.5078},
 primaryClass = {astro-ph.GA},
       adsurl = {https://ui.adsabs.harvard.edu/abs/2014MNRAS.443.1555U}
}

@ARTICLE{Elia2025,
       author = {{Elia}, Davide and {Evans}, II, Neal J. and {Soler}, Juan D. and {Strafella}, Francesco and {Schisano}, Eugenio and {Molinari}, Sergio and {Giannetti}, Andrea and {Patra}, Sudeshna},
        title = "{Measuring Star Formation Rates in the Milky Way from Hi-GAL 70 {\ensuremath{\mu}}m Observations}",
      journal = {\apj},
         year = 2025,
        month = feb,
       volume = {980},
       number = {2},
          eid = {216},
        pages = {216},
          doi = {10.3847/1538-4357/adaeb2},
archivePrefix = {arXiv},
       eprint = {2501.14471},
 primaryClass = {astro-ph.GA},
       adsurl = {https://ui.adsabs.harvard.edu/abs/2025ApJ...980..216E}
}

@ARTICLE{Eisenstein1998,
       author = {{Eisenstein}, Daniel J. and {Hut}, Piet},
        title = "{HOP: A New Group-Finding Algorithm for N-Body Simulations}",
      journal = {\apj},
         year = 1998,
        month = may,
       volume = {498},
       number = {1},
        pages = {137-142},
          doi = {10.1086/305535},
archivePrefix = {arXiv},
       eprint = {astro-ph/9712200},
 primaryClass = {astro-ph},
       adsurl = {https://ui.adsabs.harvard.edu/abs/1998ApJ...498..137E}
}

@ARTICLE{Colman2024,
       author = {{Colman}, Tine and {Brucy}, No{\'e} and {Girichidis}, Philipp and {Glover}, Simon C.~O. and {Benedettini}, Milena and {Soler}, Juan D. and {Tress}, Robin G. and {Traficante}, Alessio and {Hennebelle}, Patrick and {Klessen}, Ralf S. and {Molinari}, Sergio and {Miville-Desch{\^e}nes}, Marc-Antoine},
        title = "{Cloud properties across spatial scales in simulations of the interstellar medium}",
      journal = {\aap},
         year = 2024,
        month = jun,
       volume = {686},
          eid = {A155},
        pages = {A155},
          doi = {10.1051/0004-6361/202348983},
archivePrefix = {arXiv},
       eprint = {2403.00512},
 primaryClass = {astro-ph.GA},
       adsurl = {https://ui.adsabs.harvard.edu/abs/2024A&A...686A.155C}
}

@ARTICLE{Elia2021,
       author = {{Elia}, Davide and {Merello}, M. and {Molinari}, S. and {Schisano}, E. and {Zavagno}, A. and {Russeil}, D. and {M{\`e}ge}, P. and {Martin}, P.~G. and {Olmi}, L. and {Pestalozzi}, M. and {Plume}, R. and {Ragan}, S.~E. and {Benedettini}, M. and {Eden}, D.~J. and {Moore}, T.~J.~T. and {Noriega-Crespo}, A. and {Paladini}, R. and {Palmeirim}, P. and {Pezzuto}, S. and {Pilbratt}, G.~L. and {Rygl}, K.~L.~J. and {Schilke}, P. and {Strafella}, F. and {Tan}, J.~C. and {Traficante}, A. and {Baldeschi}, A. and {Bally}, J. and {di Giorgio}, A.~M. and {Fiorellino}, E. and {Liu}, S.~J. and {Piazzo}, L. and {Polychroni}, D.},
        title = "{The Hi-GAL compact source catalogue - II. The 360{\textdegree} catalogue of clump physical properties}",
      journal = {\mnras},
         year = 2021,
        month = jun,
       volume = {504},
       number = {2},
        pages = {2742-2766},
          doi = {10.1093/mnras/stab1038},
archivePrefix = {arXiv},
       eprint = {2104.04807},
 primaryClass = {astro-ph.GA},
       adsurl = {https://ui.adsabs.harvard.edu/abs/2021MNRAS.504.2742E}
}

@ARTICLE{Schmidt1959,
       author = {{Schmidt}, Maarten},
        title = "{The Rate of Star Formation.}",
      journal = {\apj},
         year = 1959,
        month = mar,
       volume = {129},
        pages = {243},
          doi = {10.1086/146614},
       adsurl = {https://ui.adsabs.harvard.edu/abs/1959ApJ...129..243S}
}

@ARTICLE{Kennicutt1989,
       author = {{Kennicutt}, Jr., Robert C.},
        title = "{The Star Formation Law in Galactic Disks}",
      journal = {\apj},
         year = 1989,
        month = sep,
       volume = {344},
        pages = {685},
          doi = {10.1086/167834},
       adsurl = {https://ui.adsabs.harvard.edu/abs/1989ApJ...344..685K}
}

@ARTICLE{Molinari2016,
       author = {{Molinari}, S. and {Schisano}, E. and {Elia}, D. and {Pestalozzi}, M. and {Traficante}, A. and {Pezzuto}, S. and {Swinyard}, B.~M. and {Noriega-Crespo}, A. and {Bally}, J. and {Moore}, T.~J.~T. and {Plume}, R. and {Zavagno}, A. and {di Giorgio}, A.~M. and {Liu}, S.~J. and {Pilbratt}, G.~L. and {Mottram}, J.~C. and {Russeil}, D. and {Piazzo}, L. and {Veneziani}, M. and {Benedettini}, M. and {Calzoletti}, L. and {Faustini}, F. and {Natoli}, P. and {Piacentini}, F. and {Merello}, M. and {Palmese}, A. and {Del Grande}, R. and {Polychroni}, D. and {Rygl}, K.~L.~J. and {Polenta}, G. and {Barlow}, M.~J. and {Bernard}, J.-P. and {Martin}, P.~G. and {Testi}, L. and {Ali}, B. and {Andr{\'e}}, P. and {Beltr{\'a}n}, M.~T. and {Billot}, N. and {Carey}, S. and {Cesaroni}, R. and {Compi{\`e}gne}, M. and {Eden}, D. and {Fukui}, Y. and {Garcia-Lario}, P. and {Hoare}, M.~G. and {Huang}, M. and {Joncas}, G. and {Lim}, T.~L. and {Lord}, S.~D. and {Martinavarro-Armengol}, S. and {Motte}, F. and {Paladini}, R. and {Paradis}, D. and {Peretto}, N. and {Robitaille}, T. and {Schilke}, P. and {Schneider}, N. and {Schulz}, B. and {Sibthorpe}, B. and {Strafella}, F. and {Thompson}, M.~A. and {Umana}, G. and {Ward-Thompson}, D. and {Wyrowski}, F.},
        title = "{Hi-GAL, the Herschel infrared Galactic Plane Survey: photometric maps and compact source catalogues. First data release for the inner Milky Way: +68{\textdegree} {\ensuremath{\geq}} l {\ensuremath{\geq}} -70{\textdegree}}",
      journal = {\aap},
         year = 2016,
        month = jul,
       volume = {591},
          eid = {A149},
        pages = {A149},
          doi = {10.1051/0004-6361/201526380},
archivePrefix = {arXiv},
       eprint = {1604.05911},
 primaryClass = {astro-ph.GA},
       adsurl = {https://ui.adsabs.harvard.edu/abs/2016A&A...591A.149M}
}

@ARTICLE{Elia2017,
       author = {{Elia}, Davide and {Molinari}, S. and {Schisano}, E. and {Pestalozzi}, M. and {Pezzuto}, S. and {Merello}, M. and {Noriega-Crespo}, A. and {Moore}, T.~J.~T. and {Russeil}, D. and {Mottram}, J.~C. and {Paladini}, R. and {Strafella}, F. and {Benedettini}, M. and {Bernard}, J.~P. and {Di Giorgio}, A. and {Eden}, D.~J. and {Fukui}, Y. and {Plume}, R. and {Bally}, J. and {Martin}, P.~G. and {Ragan}, S.~E. and {Jaffa}, S.~E. and {Motte}, F. and {Olmi}, L. and {Schneider}, N. and {Testi}, L. and {Wyrowski}, F. and {Zavagno}, A. and {Calzoletti}, L. and {Faustini}, F. and {Natoli}, P. and {Palmeirim}, P. and {Piacentini}, F. and {Piazzo}, L. and {Pilbratt}, G.~L. and {Polychroni}, D. and {Baldeschi}, A. and {Beltr{\'a}n}, M.~T. and {Billot}, N. and {Cambr{\'e}sy}, L. and {Cesaroni}, R. and {Garc{\'\i}a-Lario}, P. and {Hoare}, M.~G. and {Huang}, M. and {Joncas}, G. and {Liu}, S.~J. and {Maiolo}, B.~M.~T. and {Marsh}, K.~A. and {Maruccia}, Y. and {M{\`e}ge}, P. and {Peretto}, N. and {Rygl}, K.~L.~J. and {Schilke}, P. and {Thompson}, M.~A. and {Traficante}, A. and {Umana}, G. and {Veneziani}, M. and {Ward-Thompson}, D. and {Whitworth}, A.~P. and {Arab}, H. and {Bandieramonte}, M. and {Becciani}, U. and {Brescia}, M. and {Buemi}, C. and {Bufano}, F. and {Butora}, R. and {Cavuoti}, S. and {Costa}, A. and {Fiorellino}, E. and {Hajnal}, A. and {Hayakawa}, T. and {Kacsuk}, P. and {Leto}, P. and {Li Causi}, G. and {Marchili}, N. and {Martinavarro-Armengol}, S. and {Mercurio}, A. and {Molinaro}, M. and {Riccio}, G. and {Sano}, H. and {Sciacca}, E. and {Tachihara}, K. and {Torii}, K. and {Trigilio}, C. and {Vitello}, F. and {Yamamoto}, H.},
        title = "{The Hi-GAL compact source catalogue - I. The physical properties of the clumps in the inner Galaxy (-71.0{\textdegree} < {\ensuremath{\ell}} < 67.0{\textdegree})}",
      journal = {\mnras},
         year = 2017,
        month = oct,
       volume = {471},
       number = {1},
        pages = {100-143},
          doi = {10.1093/mnras/stx1357},
archivePrefix = {arXiv},
       eprint = {1706.01046},
 primaryClass = {astro-ph.GA},
       adsurl = {https://ui.adsabs.harvard.edu/abs/2017MNRAS.471..100E}
}

@ARTICLE{Elia2022,
       author = {{Elia}, D. and {Molinari}, S. and {Schisano}, E. and {Soler}, J.~D. and {Merello}, M. and {Russeil}, D. and {Veneziani}, M. and {Zavagno}, A. and {Noriega-Crespo}, A. and {Olmi}, L. and {Benedettini}, M. and {Hennebelle}, P. and {Klessen}, R.~S. and {Leurini}, S. and {Paladini}, R. and {Pezzuto}, S. and {Traficante}, A. and {Eden}, D.~J. and {Martin}, P.~G. and {Sormani}, M. and {Coletta}, A. and {Colman}, T. and {Plume}, R. and {Maruccia}, Y. and {Mininni}, C. and {Liu}, S.~J.},
        title = "{The Star Formation Rate of the Milky Way as Seen by Herschel}",
      journal = {\apj},
         year = 2022,
        month = dec,
       volume = {941},
       number = {2},
          eid = {162},
        pages = {162},
          doi = {10.3847/1538-4357/aca27d},
archivePrefix = {arXiv},
       eprint = {2211.05573},
 primaryClass = {astro-ph.GA},
       adsurl = {https://ui.adsabs.harvard.edu/abs/2022ApJ...941..162E}
}

@ARTICLE{Colling2018,
       author = {{Colling}, C{\'e}dric and {Hennebelle}, Patrick and {Geen}, Sam and {Iffrig}, Olivier and {Bournaud}, Fr{\'e}d{\'e}ric},
        title = "{Impact of galactic shear and stellar feedback on star formation}",
      journal = {\aap},
         year = 2018,
        month = dec,
       volume = {620},
          eid = {A21},
        pages = {A21},
          doi = {10.1051/0004-6361/201833161},
archivePrefix = {arXiv},
       eprint = {1809.01037},
 primaryClass = {astro-ph.GA},
       adsurl = {https://ui.adsabs.harvard.edu/abs/2018A&A...620A..21C}
}

@ARTICLE{Verliat2022,
       author = {{Verliat}, Antoine and {Hennebelle}, Patrick and {Gonz{\'a}lez}, Marta and {Lee}, Yueh-Ning and {Geen}, Sam},
        title = "{Influence of protostellar jets and HII regions on the formation and evolution of stellar clusters}",
      journal = {\aap},
         year = 2022,
        month = jul,
       volume = {663},
          eid = {A6},
        pages = {A6},
          doi = {10.1051/0004-6361/202141765},
archivePrefix = {arXiv},
       eprint = {2202.02237},
 primaryClass = {astro-ph.GA},
       adsurl = {https://ui.adsabs.harvard.edu/abs/2022A&A...663A...6V}
}

@ARTICLE{Larson1981,
       author = {{Larson}, R.~B.},
        title = "{Turbulence and star formation in molecular clouds.}",
      journal = {\mnras},
         year = 1981,
        month = mar,
       volume = {194},
        pages = {809-826},
          doi = {10.1093/mnras/194.4.809},
       adsurl = {https://ui.adsabs.harvard.edu/abs/1981MNRAS.194..809L}
}

@ARTICLE{Lenkic2024,
       author = {{Lenki{\'c}}, Laura and {Nally}, Conor and {Jones}, Olivia C. and {Boyer}, Martha L. and {Kavanagh}, Patrick J. and {Habel}, Nolan and {Nayak}, Omnarayani and {Hirschauer}, Alec S. and {Meixner}, Margaret and {Sargent}, B.~A. and {Temim}, Tea},
        title = "{A JWST/MIRI and NIRCam Analysis of the Young Stellar Object Population in the Spitzer I Region of NGC 6822}",
      journal = {\apj},
         year = 2024,
        month = jun,
       volume = {967},
       number = {2},
          eid = {110},
        pages = {110},
          doi = {10.3847/1538-4357/ad3f90},
archivePrefix = {arXiv},
       eprint = {2307.15704},
 primaryClass = {astro-ph.SR},
       adsurl = {https://ui.adsabs.harvard.edu/abs/2024ApJ...967..110L}
}

@ARTICLE{Crompvoets2024,
       author = {{Crompvoets}, B.~L. and {Di Francesco}, J. and {Teimoorinia}, H. and {Preibisch}, T.},
        title = "{Climbing the Cliffs: Classifying Young Stellar Objects in the Cosmic Cliffs JWST Data Using a Probabilistic Random Forest}",
      journal = {\aj},
         year = 2024,
        month = aug,
       volume = {168},
       number = {2},
          eid = {63},
        pages = {63},
          doi = {10.3847/1538-3881/ad51fc},
archivePrefix = {arXiv},
       eprint = {2301.04772},
 primaryClass = {astro-ph.SR},
       adsurl = {https://ui.adsabs.harvard.edu/abs/2024AJ....168...63C}
}

@ARTICLE{Hennebelle2018,
       author = {{Hennebelle}, Patrick},
        title = "{The FRIGG project: From intermediate galactic scales to self-gravitating cores}",
      journal = {\aap},
         year = 2018,
        month = mar,
       volume = {611},
          eid = {A24},
        pages = {A24},
          doi = {10.1051/0004-6361/201731071},
archivePrefix = {arXiv},
       eprint = {1709.07619},
 primaryClass = {astro-ph.GA},
       adsurl = {https://ui.adsabs.harvard.edu/abs/2018A&A...611A..24H}
}

@ARTICLE{Nakamura2005,
       author = {{Nakamura}, Fumitaka and {Li}, Zhi-Yun},
        title = "{Quiescent Cores and the Efficiency of Turbulence-accelerated, Magnetically Regulated Star Formation}",
      journal = {\apj},
         year = 2005,
        month = sep,
       volume = {631},
       number = {1},
        pages = {411-428},
          doi = {10.1086/432606},
archivePrefix = {arXiv},
       eprint = {astro-ph/0502130},
 primaryClass = {astro-ph},
       adsurl = {https://ui.adsabs.harvard.edu/abs/2005ApJ...631..411N}
}

@ARTICLE{Bate1995,
       author = {{Bate}, Matthew R. and {Bonnell}, Ian A. and {Price}, Nigel M.},
        title = "{Modelling accretion in protobinary systems}",
      journal = {\mnras},
         year = 1995,
        month = nov,
       volume = {277},
       number = {2},
        pages = {362-376},
          doi = {10.1093/mnras/277.2.362},
archivePrefix = {arXiv},
       eprint = {astro-ph/9510149},
 primaryClass = {astro-ph},
       adsurl = {https://ui.adsabs.harvard.edu/abs/1995MNRAS.277..362B}
}

@ARTICLE{Clark2014,
       author = {{Clark}, Paul C. and {Glover}, Simon C.~O.},
        title = "{On column density thresholds and the star formation rate}",
      journal = {\mnras},
         year = 2014,
        month = nov,
       volume = {444},
       number = {3},
        pages = {2396-2414},
          doi = {10.1093/mnras/stu1589},
archivePrefix = {arXiv},
       eprint = {1306.5714},
 primaryClass = {astro-ph.GA},
       adsurl = {https://ui.adsabs.harvard.edu/abs/2014MNRAS.444.2396C}
}

@ARTICLE{Grudic2019,
       author = {{Grudi{\'c}}, Michael Y. and {Hopkins}, Philip F. and {Lee}, Eve J. and {Murray}, Norman and {Faucher-Gigu{\`e}re}, Claude-Andr{\'e} and {Johnson}, L. Clifton},
        title = "{On the nature of variations in the measured star formation efficiency of molecular clouds}",
      journal = {\mnras},
         year = 2019,
        month = sep,
       volume = {488},
       number = {2},
        pages = {1501-1518},
          doi = {10.1093/mnras/stz1758},
archivePrefix = {arXiv},
       eprint = {1809.08348},
 primaryClass = {astro-ph.GA},
       adsurl = {https://ui.adsabs.harvard.edu/abs/2019MNRAS.488.1501G}
}

@ARTICLE{Federrath2010,
       author = {{Federrath}, Christoph and {Banerjee}, Robi and {Clark}, Paul C. and {Klessen}, Ralf S.},
        title = "{Modeling Collapse and Accretion in Turbulent Gas Clouds: Implementation and Comparison of Sink Particles in AMR and SPH}",
      journal = {\apj},
         year = 2010,
        month = apr,
       volume = {713},
       number = {1},
        pages = {269-290},
          doi = {10.1088/0004-637X/713/1/269},
archivePrefix = {arXiv},
       eprint = {1001.4456},
 primaryClass = {astro-ph.SR},
       adsurl = {https://ui.adsabs.harvard.edu/abs/2010ApJ...713..269F}
}

@ARTICLE{Heyer2004,
       author = {{Heyer}, Mark H. and {Brunt}, Christopher M.},
        title = "{The Universality of Turbulence in Galactic Molecular Clouds}",
      journal = {\apjl},
         year = 2004,
        month = nov,
       volume = {615},
       number = {1},
        pages = {L45-L48},
          doi = {10.1086/425978},
archivePrefix = {arXiv},
       eprint = {astro-ph/0409420},
 primaryClass = {astro-ph},
       adsurl = {https://ui.adsabs.harvard.edu/abs/2004ApJ...615L..45H}
}

@ARTICLE{Solomon1987,
       author = {{Solomon}, P.~M. and {Rivolo}, A.~R. and {Barrett}, J. and {Yahil}, A.},
        title = "{Mass, Luminosity, and Line Width Relations of Galactic Molecular Clouds}",
      journal = {\apj},
         year = 1987,
        month = aug,
       volume = {319},
        pages = {730},
          doi = {10.1086/165493},
       adsurl = {https://ui.adsabs.harvard.edu/abs/1987ApJ...319..730S}
}

@ARTICLE{Salpeter1955,
       author = {{Salpeter}, Edwin E.},
        title = "{The Luminosity Function and Stellar Evolution.}",
      journal = {\apj},
         year = 1955,
        month = jan,
       volume = {121},
        pages = {161},
          doi = {10.1086/145971},
       adsurl = {https://ui.adsabs.harvard.edu/abs/1955ApJ...121..161S}
}

@ARTICLE{Veneziani2017,
       author = {{Veneziani}, M. and {Schisano}, E. and {Elia}, D. and {Noriega-Crespo}, A. and {Carey}, S. and {Di Giorgio}, A. and {Fukui}, Y. and {Maiolo}, B.~M.~T. and {Maruccia}, Y. and {Mizuno}, A. and {Mizuno}, N. and {Molinari}, S. and {Mottram}, J.~C. and {Moore}, T.~J.~T. and {Onishi}, T. and {Paladini}, R. and {Paradis}, D. and {Pestalozzi}, M. and {Pezzuto}, S. and {Piacentini}, F. and {Plume}, R. and {Russeil}, D. and {Strafella}, F.},
        title = "{An analysis of star formation with Herschel in the Hi-GAL Survey. II. The tips of the Galactic bar}",
      journal = {\aap},
         year = 2017,
        month = mar,
       volume = {599},
          eid = {A7},
        pages = {A7},
          doi = {10.1051/0004-6361/201423474},
archivePrefix = {arXiv},
       eprint = {1612.04995},
 primaryClass = {astro-ph.GA},
       adsurl = {https://ui.adsabs.harvard.edu/abs/2017A&A...599A...7V}
}

@ARTICLE{Krumholz2005,
       author = {{Krumholz}, Mark R. and {McKee}, Christopher F.},
        title = "{A General Theory of Turbulence-regulated Star Formation, from Spirals to Ultraluminous Infrared Galaxies}",
      journal = {\apj},
         year = 2005,
        month = sep,
       volume = {630},
       number = {1},
        pages = {250-268},
          doi = {10.1086/431734},
archivePrefix = {arXiv},
       eprint = {astro-ph/0505177},
 primaryClass = {astro-ph},
       adsurl = {https://ui.adsabs.harvard.edu/abs/2005ApJ...630..250K}
}

@ARTICLE{Molinari2008,
       author = {{Molinari}, S. and {Pezzuto}, S. and {Cesaroni}, R. and {Brand}, J. and {Faustini}, F. and {Testi}, L.},
        title = "{The evolution of the spectral energy distribution in massive young stellar objects}",
      journal = {\aap},
         year = 2008,
        month = apr,
       volume = {481},
       number = {2},
        pages = {345-365},
          doi = {10.1051/0004-6361:20078661},
       adsurl = {https://ui.adsabs.harvard.edu/abs/2008A&A...481..345M}
}

@ARTICLE{Hu2022,
       author = {{Hu}, Zipeng and {Krumholz}, Mark R. and {Pokhrel}, Riwaj and {Gutermuth}, Robert A.},
        title = "{High-precision star-formation efficiency measurements in nearby clouds}",
      journal = {\mnras},
         year = 2022,
        month = feb,
       volume = {511},
       number = {1},
        pages = {1431-1438},
          doi = {10.1093/mnras/stac174},
archivePrefix = {arXiv},
       eprint = {2109.04665},
 primaryClass = {astro-ph.GA},
       adsurl = {https://ui.adsabs.harvard.edu/abs/2022MNRAS.511.1431H}
}

@ARTICLE{Lee2016,
       author = {{Lee}, Yueh-Ning and {Hennebelle}, Patrick},
        title = "{Formation of a protocluster: A virialized structure from gravoturbulent collapse. II. A two-dimensional analytical model for a rotating and accreting system}",
      journal = {\aap},
         year = 2016,
        month = jun,
       volume = {591},
          eid = {A31},
        pages = {A31},
          doi = {10.1051/0004-6361/201527982},
archivePrefix = {arXiv},
       eprint = {1603.07983},
 primaryClass = {astro-ph.SR},
       adsurl = {https://ui.adsabs.harvard.edu/abs/2016A&A...591A..31L}
}

@ARTICLE{Hennebelle2012,
       author = {{Hennebelle}, P.},
        title = "{Formation of proto-clusters and star formation within clusters: apparent universality of the initial mass function?}",
      journal = {\aap},
         year = 2012,
        month = sep,
       volume = {545},
          eid = {A147},
        pages = {A147},
          doi = {10.1051/0004-6361/201219440},
archivePrefix = {arXiv},
       eprint = {1207.6776},
 primaryClass = {astro-ph.GA},
       adsurl = {https://ui.adsabs.harvard.edu/abs/2012A&A...545A.147H}
}

\clearpage
\begin{appendix}
\section{\texttt{HOP} parameter space}
\label{app: HOP parameter space}
\begin{figure}[h!]
    \centering
    \includegraphics[width=1\linewidth]{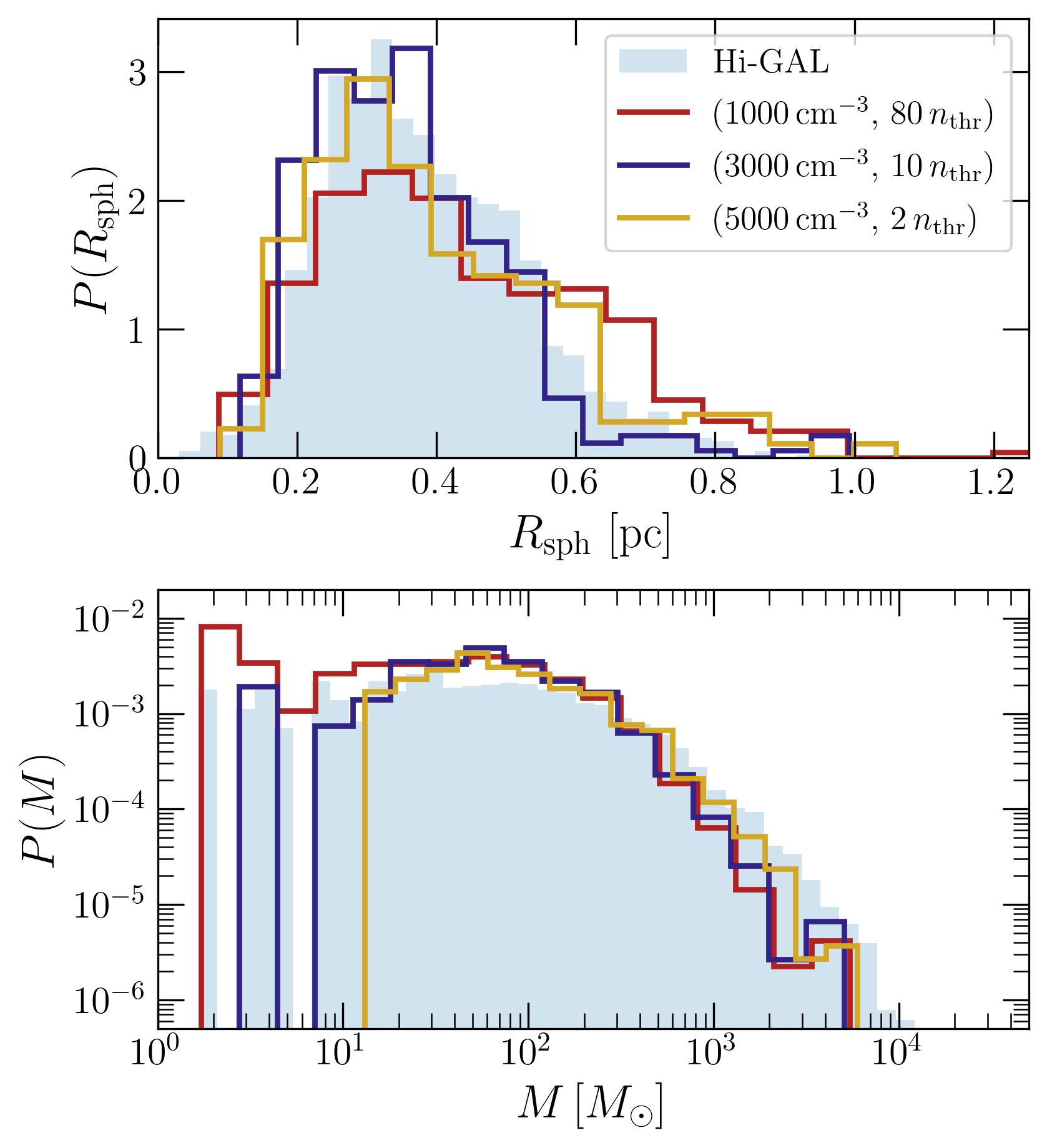}
    \caption{Joint distributions of the size (top) and mass (bottom) of the simulated clumps recovered with different \texttt{HOP} parameters: ($1000\,\ndens$, $80\,n_\mathrm{thr}$) in red;  ($3000\,\ndens$, $10\,n_\mathrm{thr}$), adopted in the text, in blue; and  ($5000\,\ndens$, $2\,n_\mathrm{thr}$) in yellow. The filled histogram shows the Hi-GAL clumps distributions.}
    \label{app: fig: RM multiple saddles}
\end{figure}

\begin{figure}[t!]
    \centering
    \includegraphics[width=1\linewidth]{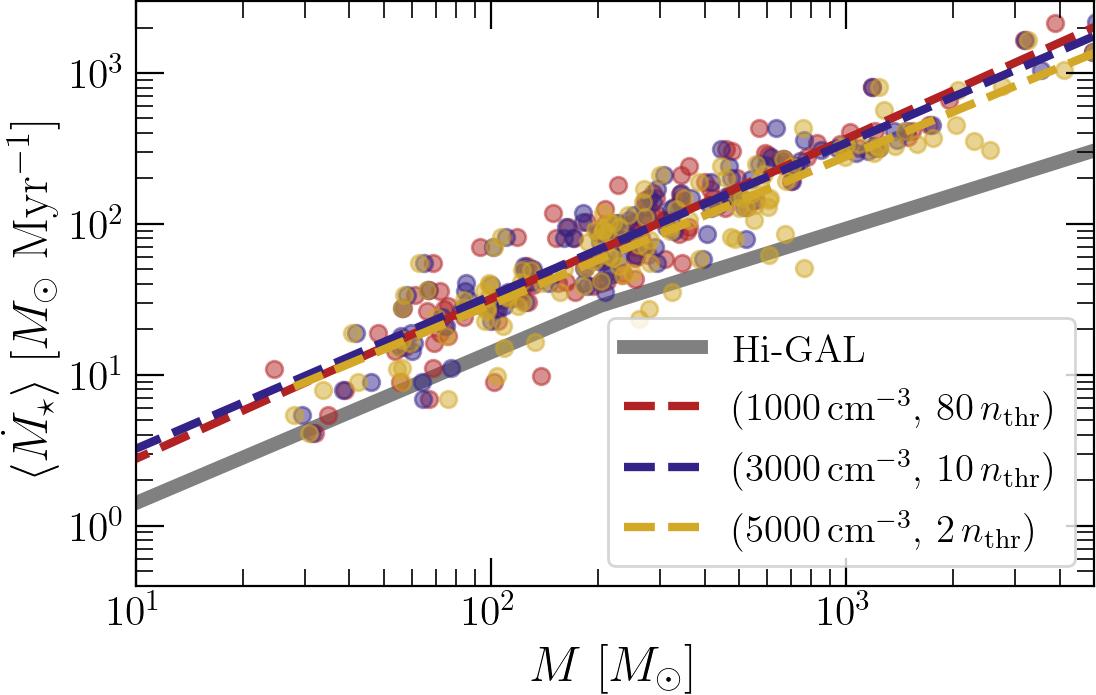}
    \caption{Relationship between the averaged $\simuSFR$ and mass of the structures obtained using different sets of \texttt{HOP} parameters (as in Fig.~\ref{app: fig: RM multiple saddles}). Only \texttt{HOP} clumps with $\tau_\star>\tff(n_\mathrm{thr})$ are retained in each panels.}
    \label{fig: app: SFR-M multiple saddles}
\end{figure}

To understand the impact of the \texttt{HOP} parameters choice on the results drawn in the text, we repeat the analysis adopting different combinations of $n_\mathrm{thr}$ and $n_\mathrm{saddle}$. Trying to match the average size of the Hi-GAL clumps, we find other two sets of pairs ($n_\mathrm{thr}$, $n_\mathrm{saddle}$):  ($1000\,\ndens$, $80\,n_\mathrm{thr}$) and ($5000\,\ndens$, $2\,n_\mathrm{thr}$). The joint distributions arising in these two cases, along with the one adopted in the text, are shown in Fig.~\ref{app: fig: RM multiple saddles}. Despite a few differences may be noticed in the tail, the bulk of the distribution is preserved, with the peak position varying only slightly. Even the plateau and the high-mass tail of the mass distribution remains roughly conserved, despite some deviations likely due to completeness.
Nonetheless, what proves the analysis to be robust is the resilience of the $M-\simuSFR$ relation, shown in Fig.~\ref{fig: app: SFR-M multiple saddles}. The joint fit of \texttt{HOP} clumps obtained with the different parameters overlap each other, proving that the conclusions drawn on the SF process do not depend on the choice of the parameter set. 

\section{Temporal evolution of clump properties}
\label{app: temporal evolution}
\begin{figure}
    \centering
    \includegraphics[width=1\linewidth]{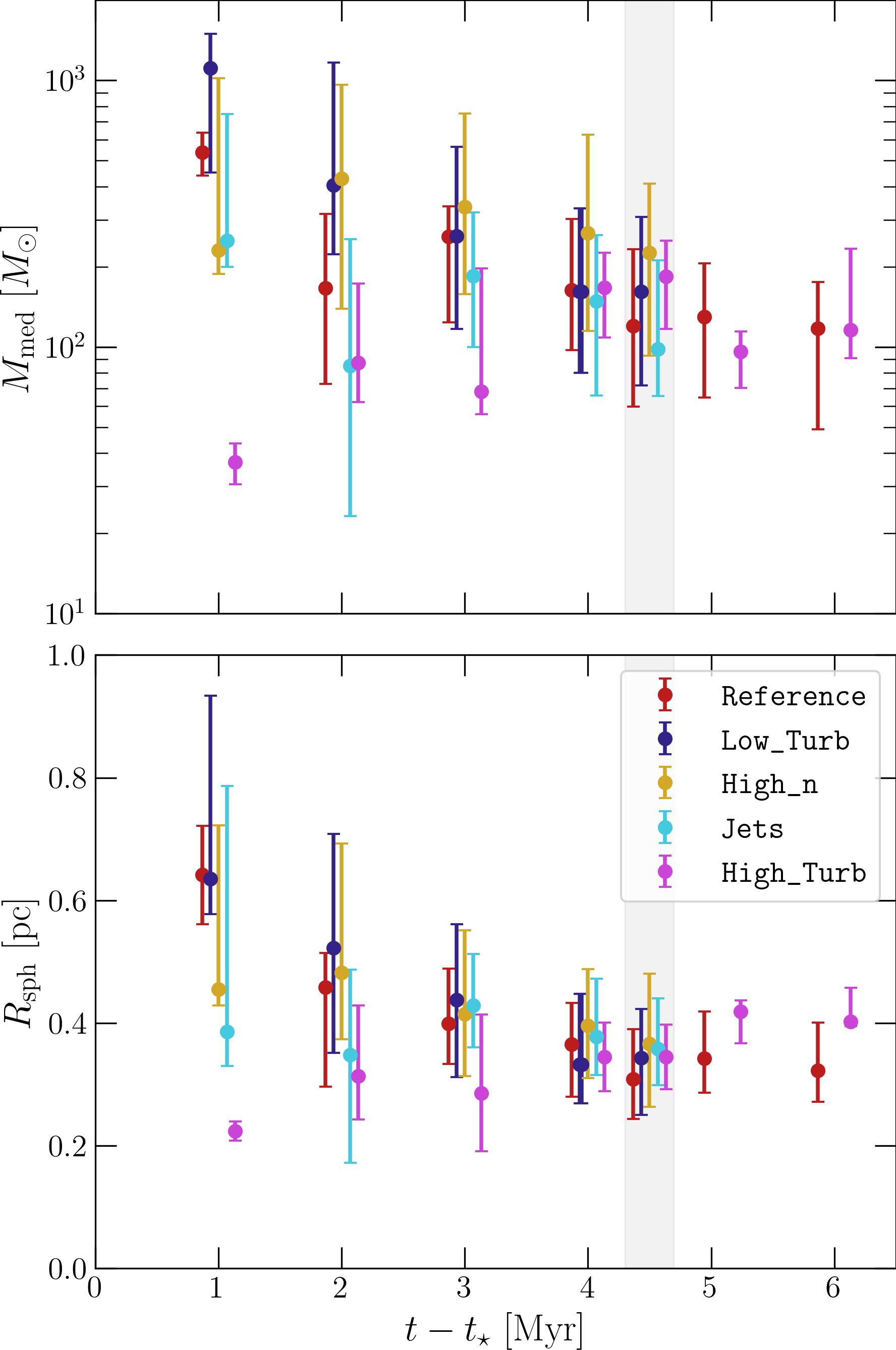}
    \caption{Temporal evolution of the \texttt{HOP} clump mass (top) and radius (bottom) across the various simulations (red: \refsim; blue: \lowturb; yellow: \highn; purple: \highturb; cyan: \outflowsim). The points mark the distribution's meadian, while the error bars mark the 25th and 75th percentiles. The snapshots are taken at the same time, which have been slightly displaced for clarity. The shaded grey area highlights the reference snapshot selected for the analysis performed in the text.}
    \label{fig: temporal evolution properties}
\end{figure}

One important check is whether the population had reached a stable distribution by the time we analyse the sample. Figure~\ref{fig: temporal evolution properties} shows the temporal evolution of the mean mass (top) and spherical radius (bottom) of the star-forming \texttt{HOP} clumps, as a function of time elapsed since the onset of SF ($t_\star$). Error bars indicate the asymmetric $1\sigma$ dispersion of each distribution. During the first $3\,\Myr$, all simulations exhibit a clear trend towards relaxation. Both mass and size of the structures decrease, likely due to progressive fragmentation and collapse as the clumps evolve. At later times, the distributions start to behave differently.
In particular, the average clump mass in \highn does not fully reach a plateau, with slight variations also being detectable in the average size. However, between $3$ and $4.5\,\Myr$ the mass variation is limited to a factor of $\lesssim 1.5$ and remains within the intrinsic dispersion. The distributions from \highturb also show strong fluctuations, but in this case these are probably driven by limited statistics (typically 3--9 star-forming clumps per snapshot). In contrast, \refsim, \lowturb, and \outflowsim simulations seem to have attained a certain stability around the selected snapshot. Extending the analysis to later times in \refsim, the distribution seems to remain quite stable.
Overall, these trends indicate that the population at the selected snapshot is quite representative of that generated by the large-scale properties of the kiloparsec domain. The remaining temporal evolution of the distributions appears, overall, less prominent than the intrinsic dispersion of the structures' properties. 

\section{The mass--SFR relation at different scales}
\label{app: HOP thresholds}
In the text, we selected the \texttt{HOP} parameter to match the properties of Hi-GAL clumps. However, as discussed in Section~\ref{sec: what is a clump}, the SF process within this structure is more related to the scale analysed, and whether or not clumps can be regarded as a real structure separated from the ambient medium ceases to be significant. To verify this statement, we can explore the behaviour of the structure retrieved at different $n_\mathrm{thr}$. Lowering (or increasing) the threshold, the recovered \texttt{HOP} structures will include larger (or smaller) portions of the ambient medium. If a net transition between the clump and the background gas does not exist, the scaling relation will remain preserved even at different $n_\mathrm{thr}$ (keeping $n_\mathrm{saddle}=2\,n_\mathrm{saddle}$).

\begin{figure*}[t!]
    \centering
    \includegraphics[width=1\linewidth]{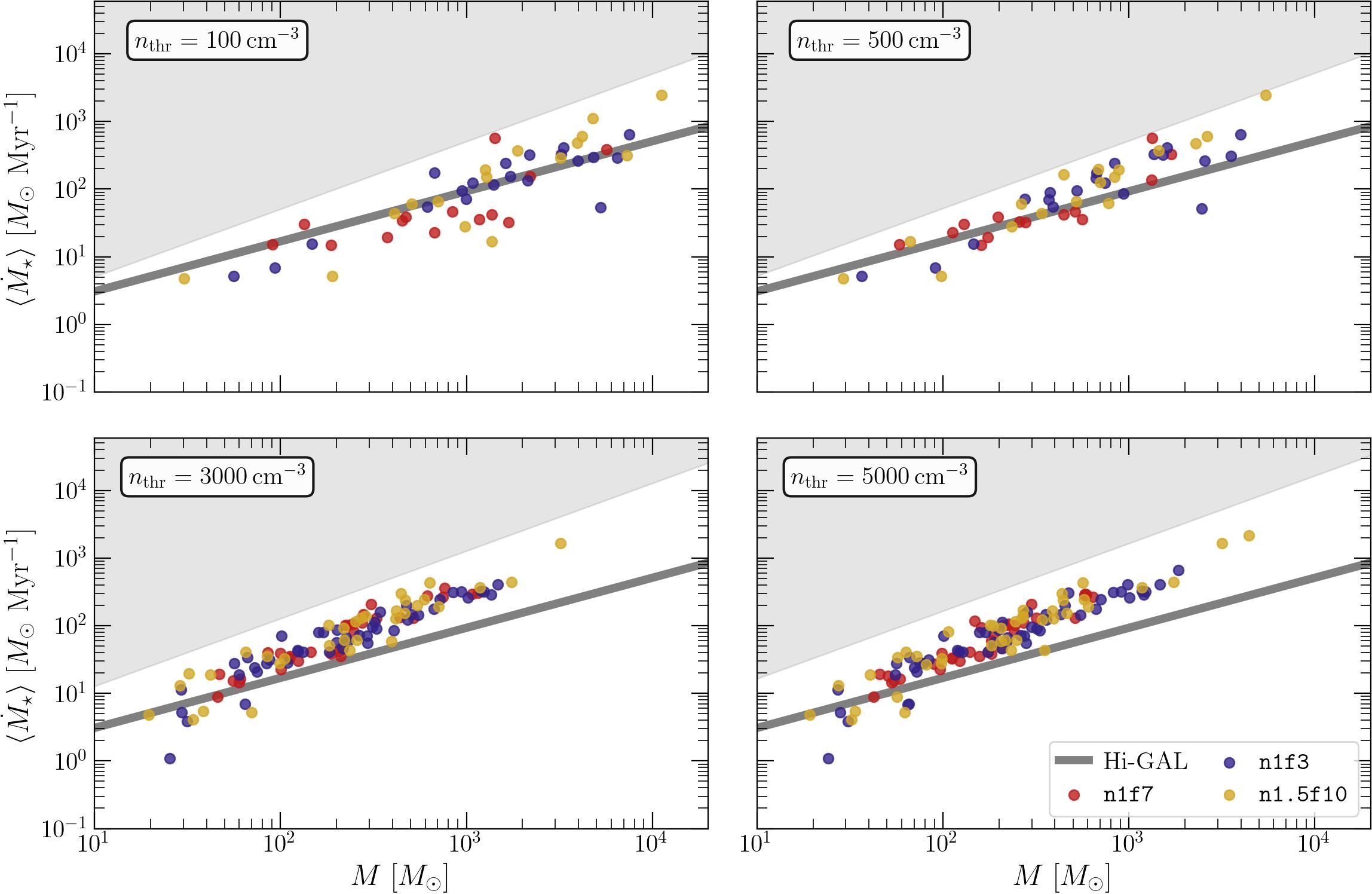}
    \caption{Relationship between the averaged $\simuSFR$ and mass of the structures retrieved at different $n_\mathrm{thr}$, preserving the scaling $n_\mathrm{peak}=2\,n_\mathrm{thr}$ and $n_\mathrm{saddle}=10\,n_\mathrm{thr}$. For clarity, we only display the simulations without outflows and with sufficient statistics.
    Only \texttt{HOP} clumps with $\tau_\star>\tff(n_\mathrm{thr})$ are retained in each panels, except for $n_\mathrm{thr}=100\,\ndens$ where $\tau_\star>\frac{1}{2}\tff(n_\mathrm{thr})$ was used. The shaded area marks the theoretically forbidden region -- $\langle\simuSFR\rangle\geq M/t_\mathrm{mask}$, with $t_\mathrm{mask}=\frac{1}{2}\tff$ in the top left panel, and $t_\mathrm{mask}=\tff$ in the others.}
    \label{fig: SFR multilpe Thr}
\end{figure*}

Figure~\ref{fig: SFR multilpe Thr} shows the $M-\simuSFR$ relation in the clumps recovered using different thresholds ($n_\mathrm{thr}=100,\,500,\,3000,\,$ and $5000\,\ndens$), while Tables~\ref{tab:sfr_alpha_by_threshold}-\ref{tab:sfr_offs_by_threshold} show the logarithmic slopes and the offsets at $M=100\,\MSun$ recovered through the power law fit ($\alpha_X$ and $\dot{M}_\star^{X}$ identifying the exponent and offset for the threshold $n_\mathrm{thr}=X\,\ndens$). In the various extractions, we retained only clumps containing a sink older than $\tff(n_\mathrm{thr})$ (as done in the text). Only in the case of $n_\mathrm{thr}=100\,\ndens$, where $\tff$ is comparable with the time elapsed since the onset of SF ($\tff\approx4.3\,\Myr$), we filtered using $\tau_\star\geq\frac{1}{2}\tff$. 

From the analyses of these new sets, we can note a few interesting features. 
First, we see a negative dependence of the $\simuSFR$ on $n_\mathrm{thr}$. This can be explained through two effects. The first is simply due to the greater amount of gas enclosed in the structure, which increases its mass while keeping the inner SFR unchanged. This naturally decreases the offset of the linear relation. For example, between the $\simuSFR$ observed at $M=100\,\MSun$ in the clumps recovered at $n_\mathrm{thr}=500\,\ndens$ and $n_\mathrm{thr}=5000\,\ndens$, we see an increase by roughly a factor of 2 (cf. Table~\ref{tab:sfr_offs_by_threshold}). The second effect comes from the merging of multiple structures. In this case, both the mass and the $\simuSFR$ of the new clump will be increased. Since the $\simuSFR$ of the structures does not seem to vary much between the panels, we conclude that the first effect is likely to dominate over the second. 

More importantly, we confirm that even at $n_\mathrm{thr}=500\,\ndens$ and  $n_\mathrm{thr}=5000\,\ndens$ all simulations settle on a shared linear relation, supporting a hierarchical picture of the SF process rather than one dominated by the identity of individual structures. Only in  $n_\mathrm{thr}=100\,\ndens$ the relation becomes noticeably shallower. While this may be due to the less stringent filtering of old clumps, we also note that the full, unfiltered sample in Fig.~\ref{fig: SFR vs M} tended to steepen the relation rather than flatten it. Another possible explanation is a transition to turbulence-dominated accretion of the clumps (see discussion in Section~\ref{sec: models on SFR}).

\begin{table*}
\centering
\renewcommand{\arraystretch}{1.2}
\setlength{\tabcolsep}{10pt}
\caption{Best-fit slopes of the mass--SFR relation
for different density thresholds.\label{tab:sfr_alpha_by_threshold}
}
\label{tab:sfr_alpha_by_threshold}
\begin{tabular}{l c c c c}
\hline\hline
Simulation & $\alpha_{100}$ & $\alpha_{500}$ & $\alpha_{3000}$ & $\alpha_{5000}$ \\
\hline\hline
\refsim   & $0.7\pm0.2$ & $0.95\pm0.15$ & $1.09\pm0.07$ & $1.17\pm0.08$ \\
\lowturb  & $0.84\pm0.09$ & $0.92\pm0.11$ & $1.03\pm0.05$ & $1.06\pm0.05$ \\
\highn    & $0.9\pm0.1$ & $1.04\pm0.11$ & $1.09\pm0.09$ & $1.10\pm0.08$ \\
\hline\hline
\end{tabular}
\vspace{0.2cm}
\tablefoot{
$\alpha_X$ denotes the logarithmic slope obtained using
the \texttt{HOP} threshold $n_\mathrm{thr}=X\,\ndens$.
Uncertainties are $1\sigma$.
}
\end{table*}

\begin{table*}
\centering
\renewcommand{\arraystretch}{1.2}
\setlength{\tabcolsep}{10pt}
\caption{Normalisations of the mass--SFR relation
for different density thresholds.}
\label{tab:sfr_offs_by_threshold}
\begin{tabular}{l c c c c}
\hline\hline\noalign{\vskip 2pt}
Simulation & $\simuSFR^{100}(M=100\,\MSun)$ & $\simuSFR^{500}(M=100\,\MSun)$ & $\simuSFR^{3000}(M=100\,\MSun)$ & $\simuSFR^{5000}(M=100\,\MSun)$ \\\noalign{\vskip 2pt}
\hline\hline
\refsim   & $12$ & $15$ & $30$ & $33$ \\
\lowturb  & $11$ & $17$ & $32$ & $36$ \\
\highn    & $11$ & $16$ & $33$ & $32$ \\
\hline\hline
\end{tabular}
\vspace{0.2cm}
\tablefoot{
Entries give the SFR at $M=100\,\MSun$, evaluated from
the fitted mass--SFR relation, in units of $\MSun\,\Myr^{-1}$.
The superscript $X$ identifies the \texttt{HOP} threshold
$n_\mathrm{thr}=X\,\ndens$.
}
\end{table*}

\section{Analytical scaling of the mass-SFR relation}
\label{app: derivation}
Here we briefly extend the derivation of Eq.~\ref{eq: accretion models with coeff}. 
Using the uniform sphere approximation and imposing the spherical geometry onto the accreting surface ($C=4\pi$), we can rewrite Eq.~\ref{eq: accretion models with coeff} as:
\begin{equation}
    \dot{M}_\mathrm{clump}\approx 4\pi\, v_\mathrm{r}\, \rho_\mathrm{thr} \left(\frac{3M}{4\pi \langle \rho \rangle}\right)^{2/3},
        \label{app: eq: starting point0}
\end{equation}
which is equivalent to:
\begin{equation}
    \dot{M}_\mathrm{clump}= \left(36\pi\right)^{1/3}v_\mathrm{r}\,\rho_\mathrm{thr} M^{2/3}\langle \rho \rangle^{-2/3},
    \label{app: eq: starting point}
\end{equation}
with $\langle \rho \rangle$ the average density of the clump. 

\subsection{Gravity}
If the accretion is mainly driven by gravity, then the infall velocity will be dictated by the central potential, so that:
\begin{equation}
    v_\mathrm{r}=f\sqrt{\frac{GM}{R}}=\left(\frac{4\pi}{3}\right)^{1/6}f\sqrt{G}\,M^{1/3}\langle \rho \rangle.
    \label{app: eq: vdisp grav}
\end{equation}
Substituting into Eq.~\ref{app: eq: starting point} we get:
\begin{equation}
    \dot{M}_\mathrm{clump}= 3.46\pi^{1/2}\,f\sqrt{G}\,M\,\rho_\mathrm{thr}^{1/2}\left(\frac{\rho_\mathrm{thr}}{\langle \rho \rangle}\right)^{1/2}.
\end{equation}
We continue $\gamma\equiv\langle \rho \rangle/\rho_\mathrm{thr}$, which is $\geq1$ by construction and in general equal to $10$ in the simulated clump populations. Substituting $\rho_\mathrm{thr}=1.4\times n_\mathrm{thr}$ (the \texttt{HOP} selected threshold, $3000\,\ndens$) and the reference values, we finally obtain
\begin{equation}
\begin{aligned}
\simuSFR(\mathrm{gravity}) \simeq
59\,\MSun\,\Myr^{-1}\,\left(\frac{f}{1/\sqrt{5}}\right)
\left(\frac{\gamma}{10}\right)^{-\frac{1}{2}}\\
&\hspace{-3.cm}\times
\left(\frac{M}{100\,\MSun}\right)
\left(\frac{n_\mathrm{thr}}{3000\,\ndens}\right)^{\frac{1}{2}}.
\end{aligned}
\end{equation}
The reference value of $f=1/\sqrt{5}$ is obtained by adopting the 1D virial velocity dispersion of a uniform sphere (Eq.~\ref{app: eq: vdisp grav}).

\subsection{Turbulence}
In this case, the derivation is more straightforward. The turbulent velocity dispersion follows the relation:
\begin{equation}
\sigma_\mathrm{1D}=\sigma_0\left(\frac{R}{1\,\pc}\right)^p,
\end{equation}
with $\sigma_0$ the normalisation constant at $1\,\pc$. 
With $v_\mathrm{in}=\sigma_\mathrm{1D}$,  Eq.~\label{app: eq: starting point0} becomes:
\begin{equation}
    \dot{M}_\mathrm{clump}\approx 4\pi\, v_\mathrm{r}\, \rho_\mathrm{thr} \left(\frac{3M}{4\pi \langle \rho \rangle}\right)^{(2+p)/3},
\end{equation}
from which is easy to derive
\begin{equation}
\begin{aligned}
\simuSFR(\mathrm{turbulence}) = 3\left(\frac{4\pi}{3}\right)^{\frac{1-p}{3}}
 \,\sigma_0\,\left(1\,\pc\right)^{-p}\,
\gamma^{-\frac{2+p}{3}}
M^{\frac{2+p}{3}}
n_\mathrm{thr}^{\frac{1-p}{3}}.
\end{aligned}
\end{equation}
Selecting the same normalisations as above, we recover the coefficient listed in the text.
\end{appendix}
\end{document}